\documentclass[11pt]{article}

\usepackage{amsfonts,amsthm}
\usepackage{amsmath,mathtools}
\usepackage{latexsym}
\usepackage{amssymb}
\usepackage{mathrsfs}
\usepackage{pifont}
\usepackage{tikz,ulem}
\usepackage[all]{xy}
\usepackage{hyperref}
\usetikzlibrary{quotes}
\usepackage{units}

\newcommand{\aw}{\mathbf{aw}(3)}
\newcommand{\haw}{\mathbf{aw}(4)}
\newcommand{\un}{\mathbb{I}}
\newcommand{\UN}{\mathbf{1}}

\newtheoremstyle{newrem}{3pt}{3pt}{}{}
{\bfseries}{.}{.5em}{}

\newtheorem{theo}{Theorem}[section]

\newtheorem*{theo*}{Theorem}
\newtheorem{lemm}[theo]{Lemma}
\newtheorem{prop}[theo]{Proposition}

\newtheorem{coro}[theo]{Corollary}

\theoremstyle{newrem}
\newtheorem{remax}[theo]{Remark}
\newenvironment{rema}
  {\pushQED{\qed}\remax}
  {\popQED\endremax}
\theoremstyle{definition}
\newtheorem{defi}[theo]{Definition}

\newtheorem*{term*}{Notation/Terminology}

\newcommand{\scq}{\textsc{q}}

\newcommand{\cA}{\mathcal{A}}
\newcommand{\cB}{\mathcal{B}}
\newcommand{\cC}{\mathcal{C}}

\newcommand{\cG}{\mathcal{G}}

\newcommand{\cP}{\mathcal{P}}

\newcommand{\cS}{\mathcal{S}}
\newcommand{\cT}{\mathcal{T}}
\newcommand{\cX}{\mathcal{X}}

          \def\fp{{\mathfrak p}}

\def\sA{\mathsf{A}}   \def\sB{\mathsf{B}}   \def\sC{\mathsf{C}}
\def\sD{\mathsf{D}}   \def\sE{\mathsf{E}}   \def\sX{\mathsf{X}}
\def\sP{\mathsf{P}}   \def\sF{\mathsf{F}}

\newcommand{\ZZ}{{\mathbb Z}}
\renewcommand{\AA}{{\mathbb A}}
\newcommand{\BB}{{\mathbb B}}
\newcommand{\PP}{{\mathbb P}}
\newcommand{\TT}{{\mathbb T}}
\newcommand{\GG}{{\mathbb G}}

\newcommand\atopn[2]{\genfrac{}{}{0pt}{}{#1}{#2}}
\newcommand{\arxiv}[1]{\href{https://arxiv.org/abs/#1}{\texttt{arXiv:#1}}}
\newcommand{\doi}[1]{\href{https://doi.org/#1}{doi:#1}}

\newcommand{\pFq}[5]{{
{}_{#1}F_{#2}\left( \genfrac{}{}{0pt}{0}{#3}{#4}\middle| #5\right)
}}
\newcommand{\pphi}[5]{{
{}_{#1}\phi_{#2}\left( \genfrac{}{}{0pt}{0}{#3}{#4}\middle| #5\right)
}}
\newcommand{\eps}{\varepsilon}

\usepackage{geometry}
\numberwithin{equation}{section}

\begin{document}
\thispagestyle{empty}
\title{\bf $q$-deformed Griffiths polynomials of Racah type\\\ }

\author{Nicolas Cramp\'e\textsuperscript{$1,2$},
Luc Frappat\textsuperscript{$2$},
Julien Gaboriaud\textsuperscript{$3$},
Eric Ragoucy\textsuperscript{$2$}
\\[.9em]
\textsuperscript{$1$}
\small Institut Denis-Poisson CNRS/UMR 7013 - Universit\'e de Tours - Universit\'e
d'Orl\'eans,\\
\small~Parc de Grandmont, 37200 Tours, France.\\[.5em]
\textsuperscript{$2$}
\small Laboratoire d'Annecy-le-Vieux de Physique Th\'eorique LAPTh,\\
\small~Universit\'e Savoie Mont Blanc, CNRS, F-74000 Annecy,
France.\\[.9em]
\textsuperscript{$3$}
\small Graduate School of Informatics, Kyoto University, Sakyo-ku, Kyoto, 606-8501,
Japan.\\[.9em]
}

\date{}
\maketitle

\bigskip\bigskip

\begin{center}
\begin{minipage}{14cm}
\begin{center}
{\bf Abstract}\\
\end{center}
New bivariate Griffiths polynomials of  $q$-Racah type are introduced and characterized.
They generalize the polynomials orthogonal on the multinomial distribution introduced by R. Griffiths fifty years ago.
They also correspond to a $q$-deformation of the Griffiths polynomials of Racah type introduced previously by the authors and collaborators. The latter are recovered from the former by a $q\to1$ limit.
We show that these new polynomials are bispectral and biorthogonal. We also exhibit some symmetry relations
that are essential in the proof of the bispectrality property.
\end{minipage}
\end{center}

\medskip

\begin{center}
\begin{minipage}{14cm}

\textbf{MSC2020 database:} 33C80; 33C45; 16G60
\end{minipage}
\end{center}

\vfill

\textbf{E-mail:} crampe1977@gmail.com, luc.frappat@lapth.cnrs.fr,
{julien.gaboriaud@umontreal.ca},\\
eric.ragoucy@lapth.cnrs.fr

\clearpage
\newpage

\tableofcontents

\section{Introduction}
In the study of special functions, a special class of polynomials, called the
($q$-)hypergeometric orthogonal polynomials, plays a special role \cite{AW, GR}.
This class comprises diverse families of polynomials who
are usually arranged in a scheme, called the ($q$-)Askey scheme,
which allows one to visualize the connections between the members of this class (see
\cite{Koek} for a review).
All these families of polynomials are orthogonal and satisfy a three-term recurrence
relation.
In addition, they satisfy a second order differential equation or a three-term
($q$-)difference equation.
They are described as bispectral, meaning that they satisfy a recurrence
relation and a difference equation in two distinct variables.
The ($q$-)Racah polynomials sit at the top of this scheme and the various other
polynomials can be obtained from the latter by limits or specializations on the
parameters.
A generalization to the multivariate case is important.
\textbf{The goal of this paper is to provide a bivariate generalization of the $q$-Racah
polynomials that possesses the bispectrality property amongst many other properties.}
Let us emphasize that the generalization introduced here is different from the one
defined in \cite{vDS}: that one is obtained as a truncation of the Koornwinder--Macdonald
multivariable generalization of the Askey--Wilson polynomials.

A number of multivariate generalizations of orthogonal polynomials have been studied
throughout the years.
In \cite{Griff, DC}, a $n$ variable generalization of the Krawtchouk polynomials (which
are a limiting case of the Racah polynomials) has been introduced through a generating
function, and it has been proven that these polynomials are orthogonal with respect to the
multinomial distribution.
They have also been obtained in the context of a probabilistic model in \cite{HR} (and
have been called Rahman polynomials for a while), expressed in terms of Gel’fand-Aomoto
hypergeometric series \cite{MT}, and appeared as the $3nj$-symbol of the oscillator
algebra \cite{Joris, KV, Zhe, CVV}.
These polynomials are usually called the Griffiths polynomials but, in this paper, we
shall call them the \textit{Griffiths polynomials of Krawtchouk type} to distinguish them
from the ones we will construct.

A $n$ variable generalization of the ($q$-)Racah polynomials has been obtained in
\cite{Trat, GR2005}; here we shall call it the \textit{Tratnik polynomials of ($q$-)Racah
type}. These multivariate polynomials are bispectral and have been studied intensively in
different contexts (see \textit{e.g.} \cite{GI,Iliev,Scar,GIV,BM,Gro}). Taking a limit on
the Tratnik polynomials of Racah type, one obtains a multivariate generalization of the
Krawtchouk polynomials, called here the \textit{Tratnik polynomials of Krawtchouk type}.
These polynomials in $n$ variables depend on $n$ parameters whereas the Griffiths
polynomials of Krawtchouk type depend on $\tfrac{1}{2}n(n+1)$ parameters. In \cite{GVZ}, a
group theoretical interpretation of the Griffiths polynomials of Krawtchouk type has been
given (see also \cite{Ros, IT, I12}). This point of view showed that the Tratnik
polynomials of Krawtchouk type can be obtained by specifying some parameters of the
Griffiths polynomials of Krawtchouk type. The latter polynomials were also expressed as
sums of a nested product of univariate Krawtchouk polynomials.

It is then natural to ask whether it is possible to define multivariate polynomials based
on univariate Racah or $q$-Racah polynomials which, in a certain limit, reproduce the
Griffiths polynomials of Krawtchouk type. Recently, studying the representation theory of
the rank 2 Racah algebra, bivariate Griffiths polynomials of Racah type have been computed
\cite{icosi} and their bispectral property has been obtained. These polynomials are
expressed as a sum of a product of three univariate Racah polynomials and share this
structure with the bivariate Griffiths polynomials of Krawtchouk type. They have
subsequently been studied in more detail in \cite{GriffithR}, where the properties of the
bivariate polynomials have been established using only the properties of the univariate
Racah polynomials.
In addition, it has been proven (see \cite{GriffithR}) that a limit of the
bivariate Griffiths polynomials of Racah type reproduces the bivariate Griffiths
polynomials of Krawtchouk type; this provides a justification for their
designation as ``bivariate Griffiths polynomials''.

In this paper, we study a $q$-deformed version of the bivariate Griffiths polynomials of
Racah type. In other words,
we study \textit{bivariate Griffiths polynomials of $q$-Racah type} which
are expressed as a sum of a product of three univariate $q$-Racah polynomials (see
relation \eqref{eq:G3p}). From these, a limit $q\to1$ leads to the Griffith polynomials of
Racah type just discussed. Using properties of the univariate $q$-Racah polynomial, we
prove numerous properties for the proposed bivariate polynomials: bispectral properties,
biorthogonality, symmetry, duality, etc. The proofs differ from the ones in
\cite{GriffithR} since the properties used there seem not to generalize to the
$q$-deformed case. Instead, we make use of a symmetry under the exchange of the two
variables, which appears to be crucial and was not yet established in \cite{GriffithR} for
the undeformed case. To establish this symmetry, we use the Racah and Biedenharn--Elliot
relations for the monovariate $q$-Racah polynomials (see relations \eqref{eq:racahrel}
and \eqref{eq:biedenrel}). These properties are proven using the
underlying algebra of the bispectral problem which is the rank 2 Askey--Wilson algebra
$\haw$. As a byproduct of this construction, we provide finite-dimensional representations
of the $\haw$ algebra.

The plan of the paper is as follows. In Section \ref{sec:Rac}, the univariate $q$-Racah
polynomials are defined in terms of $q$-hypergeometric functions
(whose definition can be found in Appendix \ref{App:def}) and their well-known
properties, such as bispectrality, duality and orthogonality, are recalled.
Supplementary relations, called contiguity relations, are proven in subsection
\ref{ssec:cont}; these are then used to prove various properties of the Tratnik
polynomials of $q$-Racah type in Section \ref{sec:Trat}.
Section \ref{sec:RBE} is devoted to the Racah relation and the Biedenharn--Elliot
relation for the univariate $q$-Racah polynomials.
The proofs use the representation theory of the rank $1$ Askey--Wilson algebra $\aw$
and rank $2$ Askey--Wilson algebra $\haw$ respectively,
whose definitions and representations are provided in Appendices \ref{app:aw3}
and \ref{app:aw4} respectively.
We then come to the main study of the paper in Section \ref{sec:Griff}: the definition
and properties of the Griffiths polynomials of $q$-Racah type. These polynomials are
written as a sum of a product of three univariate $q$-Racah polynomials. We show in
subsection \ref{ssec:symm} that these polynomials satisfy a certain symmetry under the
exchange of the two degrees and of the two variables simultaneously. Then, the duality and
biorthogonality relations are established in subsection \ref{ssec:db}. Subsection
\ref{ssec:bispectralG} contains the bispectral properties of these Griffiths polynomials.
One recurrence relation is easily proven knowing the one of the Tratnik polynomials and
the other one is deduced from it by leveraging the symmetry relation. The difference
equations are then obtained thanks to the duality relation.
We conclude the paper in Section \ref{sec:conclu} by listing some open problems.

\section{Properties of the $q$-Racah polynomials}\label{sec:Rac}
The $q$-Racah polynomials are defined (see \textit{e.g.}~\cite{Koek}) in terms of the
$q$-hypergeometric functions as follows, for $N$ a nonnegative integer and
$0\leq n \leq N$:
\begin{align}
R_n(\mu(x);\alpha,\beta,\gamma,\delta)=
\pphi{4}{3}{q^{-n},\; \alpha\beta q^{n+1},\; q^{-x},\; \gamma\delta q^{x+1}}
           {\alpha q,\; \beta\delta q,\; \gamma q}{q;q}\,,
\end{align}
where $\mu(x)=q^{-x}+\gamma\delta q^{x+1}$ and either $\alpha$, $\beta\delta$ or $\gamma$
is equal to $q^{-N-1}$. For this paper to be self-contained, the definitions and notations
pertaining to basic hypergeometric functions and $q$-Pochhammer symbols are
recalled in Appendix \ref{App:def}.

It is well-known that the $q$-Racah polynomials $R_n(\mu(x);\alpha,\beta,\gamma,\delta)$
are polynomials of degree $n$ with respect to $\mu(x)$ and possess nice properties such
as orthogonality and bispectrality. We now recall some of these properties after making a
slight change of notation which will prove useful for the rest of the paper.

Let us write the parameters $\alpha$, $\beta$, $\gamma$ and $\delta$ in the following way:
\begin{align}
\alpha=c_2\,,\qquad \beta=c_3\,,\qquad \gamma=q^{-N-1}\,,\qquad \delta=c_1c_2 q^{N+1}\,.
\end{align}
We emphasize that $\delta$ depends on $N$, which is just a choice of parameter here,
but this choice is crucial in the determination of the contiguity relations
\eqref{eq:cont-rec-rac}, \eqref{eq:cont-rec-rac2}
and for the definitions of the Tratnik polynomials \eqref{def:Trat}
or Griffiths polynomials \eqref{eq:G3p} of $q$-Racah type.

We also use a particular normalization for the $q$-Racah polynomials
$\fp_n(x;c_1,c_2,c_3;N)$ throughout this paper. For $N$ a nonnegative
integer and $0\leq n \leq N$:
\begin{align}
\label{eq:fctpn}
\fp_n(x;c_1,c_2,c_3;N) = \Omega(n;c_1,c_2,c_3;N)\,
\pphi{4}{3}{q^{-n},\; c_2c_3q^{n+1},\; q^{-x},\; c_1c_2q^{x+1}}
           {q c_2,\; c_1c_2c_3q^{N+2},\; q^{-N}}{q;q}\,,
\end{align}
where
\begin{align}\label{def:Omega}
\Omega(n;c_1,c_2,c_3;N)=
\frac{(q^{N-n+1};q)_{n}}{(q;q)_{n}}
\frac{(1-c_2c_3q^{2n+1}) (qc_2;q)_n (c_1c_2c_3q^{N+2};q)_n (qc_1;q)_{N-n}}
     {(qc_2)^{n-N/2} (c_2c_3q^{n+1};q)_{N+1} (qc_3;q)_n}\,.
\end{align}
The normalization $\Omega(n;c_1,c_2,c_3;N)$ is intimately linked to the weight and the
norm of the $q$-Racah polynomials, as will be shown below.
The function $\fp_n(x;c_1,c_2,c_3;N)$ is a polynomial of degree $n$ with respect to $q^{-x}+c_1c_2q^{x+1}$.
When $n>N$, this function is also well-defined and satisfies $\fp_n(x;c_1,c_2,c_3;N)=0$ for $x$ nonnegative integer and $x \le N$.

Lastly, note that replacing $c_i \mapsto q^{c_i}$ and taking the limit $q\to 1$,
we recover the Racah polynomials with the normalization used in \cite{GriffithR}.

\subsection{Bispectral property, duality and orthogonality}
The $q$-Racah polynomials solve a bispectral problem \textit{i.e.}~verify a recurrence
relation and a difference equation \cite{Koek}.
They are also orthogonal and satisfy a duality property. Below, these relations
are recalled in the notations used in this paper.

\paragraph{Symmetry relation.}
The $q$-Racah polynomials obey the symmetry relation
\begin{equation}\label{eq:symR}
\fp_n(x;c_1,c_2,c_3;N) = \left(\frac{c_1}{c_3}\right)^N \, \fp_n(x;c_1^{-1},c_2^{-1},c_3^{-1};N) \Big|_{q \to q^{-1}}.
\end{equation}
The notation $q \to q^{-1}$ means that the corresponding expression is defined
by replacing $q$ by $q^{-1}$ everywhere, including the Pochhammer symbols and the basic
hypergeometric functions.
This relation is a direct consequence of the relations
\begin{equation}
\Omega(n;c_1,c_2,c_3;N)\Big|_{q \to q^{-1}, c_i \to c_i^{-1}} = \left(\frac{c_3}{c_1}\right)^N \Omega(n;c_1,c_2,c_3;N),
\end{equation}
\begin{equation}
\pphi{4}{3}{q^{-n},\;c_2c_3q^{n+1}, \;q^{-x},\; c_1c_2q^{x+1} }{q c_2,\;c_1c_2c_3 q^{N+2},\;q^{-N} }{q;q}
= \pphi{4}{3}{q^{n},\;(c_2c_3q^{n+1})^{-1}, \;q^{x},\; (c_1c_2q^{x+1})^{-1} }{(qc_2)^{-1},\;(c_1c_2c_3q^{N+2})^{-1},\;q^{N} }{\dfrac{1}{q};\dfrac{1}{q}},
\end{equation}
that are themselves checked straightforwardly.

\paragraph{Duality.}
Using the invariance of the $q$-hypergeometric functions mentioned in Appendix
\ref{App:def},
we can show that the $q$-Racah polynomials satisfy the following duality relation,
for $0\leq n,x\leq N$:
\begin{align}\label{eq:dual}
\Omega(x;c_3,c_2,c_1;N) \,\fp_n(x;c_1,c_2,c_3;N)
=\Omega(n;c_1,c_2,c_3;N)\, \fp_x(n;c_3,c_2,c_1;N)\,.
\end{align}

\paragraph{Orthogonality.}
For $0\leq n,m\leq N$,
the orthogonality relation between two $q$-Racah polynomials is:
\begin{align}\label{eq:ortho}
\sum_{x=0}^N \Omega(x;c_3,c_2,c_1;N) \,\fp_n(x;c_1,c_2,c_3;N)\,\fp_m(x;c_1,c_2,c_3;N)
=\delta_{n,m}\,\Omega(n;c_1,c_2,c_3;N)\,.
\end{align}
Strictly speaking, the term \textit{orthogonality} is valid only if the weight
$\Omega(x;c_3,c_2,c_1;N)$ is strictly positive for $0\leq x \leq N$.
This condition imposes restrictions on the parameters entering in the definition of the
$q$-Racah polynomials.

The orthogonality relation composed with the duality relation leads to
\begin{align}\label{eq:ortho2}
\sum_{x=0}^N \fp_n(x;c_1,c_2,c_3;N)\,\fp_x(m;c_3,c_2,c_1;N)=\delta_{n,m}\,.
\end{align}
We call \eqref{eq:ortho2} the \textit{involution relation}.

\paragraph{Recurrence relation.}
It is well-known from the classical theory of orthogonal polynomials that
an orthogonal polynomial sequence obeys a three-term recurrence relation.
For the $q$-Racah polynomials, it is given as follows, for $0\leq n,x\leq N$:
\begin{align}\label{eq:rec-rac}
\lambda(x;c_{1}c_2)\, \fp_n(x;c_1,c_2,c_3;N)
= \sum_{\eps=0,\pm} \Phi^\eps(n+\eps;c_1,c_2,c_3;N)\, \fp_{n+\eps}(x;c_1,c_2,c_3;N)\,,
\end{align}
where
\begin{subequations}\label{eq:rec-rac-coeff}
\begin{align}
\lambda(x;c)&=-(1-q^{-x})(1-c q^{x+1})\,,
\label{eq:lambda}\\
\Phi^+(n;c_1,c_2,c_3;N)&=
\frac{-c_2q^{n-N}(1-q^n)(1-c_1q^{N-n+1})(1-c_2c_3q^{n+N+1})(1-c_3q^n)}
     {(1-c_2c_3q^{2n})(1-c_2c_3q^{2n+1})}\,,
\label{eq:rec-racC}\\
\Phi^-(n;c_1,c_2,c_3;N)&=
\frac{(1-q^{n-N})(1-c_1c_2c_3q^{n+N+2})(1-c_2q^{n+1})(1-c_2c_3q^{n+1})}
     {(1-c_2c_3q^{2n+1})(1-c_2c_3q^{2n+2})}\,,
\label{eq:rec-racA}\\
\Phi^0(n;c_1,c_2,c_3;N)&=-\Phi^+(n;c_1,c_2,c_3;N)-\Phi^-(n;c_1,c_2,c_3;N)\,.
\label{eq:a0}
\end{align}
\end{subequations}
In the above and throughout the paper, the notation $\eps=0,\pm$ should be understood
as $\eps=0,\pm1$ whenever the $\eps$ gets added or multiplied.
Let us remark that for $n=0$, the recurrence relation involves
$\fp_{-1}(x;c_1,c_2,c_3;N)$, which vanishes by convention.
For $n=N$, it involves $\fp_{N+1}(x;c_1,c_2,c_3;N)$, which is zero as mentioned above.
In the recurrence relation, the denominators appearing in the various coefficients
must not vanish. This leads to the following constraints on the parameters:
\begin{align}\label{eq:cons1}
c_2c_3 \notin \{q^{-2},q^{-3},\dots,q^{-2N}\}\,.
\end{align}
There exist apparent poles for $c_2c_3=1,q^{-1},q^{-2N-1},q^{-2N-2}$ but the residues on
these points vanish.

We also demand that the coefficients $\Phi^+(n;c_1,c_2,c_3;N)$ and
$\Phi^-(n-1;c_1,c_2,c_3;N)$ do not vanish for $n=1,\dots, N$. This leads to the
supplementary constraints:
\begin{align}\label{eq:cons3}
c_2,q\neq 0\,,\quad q,q^2,\dots,q^N \neq 1\,,\quad
c_1,c_2, c_3,c_1c_2c_3q^{N+1}\notin \{ q^{-1},q^{-2},\dots,q^{-N} \}\,.
\end{align}

\paragraph{Difference equation.}
In addition to the previous recurrence relation, the $q$-Racah polynomials also satisfy a
difference equation which can be deduced from the duality and recurrence relations.
This difference equation reads, for $0\leq n,x\leq N$:
\begin{align}\label{eq:rec-rac4}
\lambda(n;c_{2}c_3)\, \fp_n(x;c_1,c_2,c_3;N)&=
\sum_{\eps=0,\pm} \Phi^\eps(x;c_3,c_2,c_1;N)\, \fp_{n}(x-\eps;c_1,c_2,c_3;N)\,.
\end{align}
For $x=0$ (resp. $x=N$), the difference equation simplifies since
$\Phi^+(0;c_3,c_2,c_1;N)=0$ (resp. $\Phi^-(N;c_3,c_2,c_1;N)=0$) which implies that
$\fp_{n}(N+1;c_1,c_2,c_3;N)$ or $\fp_{n}(-1;c_1,c_2,c_3;N)$ never appear in the difference
equations.
As for the recurrence relation, the denominators appearing in the coefficients of the
difference equation must not vanish. This leads to the following constraints on the
parameters:
\begin{align}\label{eq:cons2}
c_1c_2 \notin \{q^{-2},\dots,q^{-2N}\}\,.
\end{align}
Due to the constraints \eqref{eq:cons3} and \eqref{eq:cons2}, $\Phi^+(x;c_3,c_2,c_1;N)$
and $\Phi^-(x-1;c_3,c_2,c_1;N)$ do not vanish for $x=1,\dots, N$.

From now on, we suppose that the constraints \eqref{eq:cons1}, \eqref{eq:cons3} and
\eqref{eq:cons2} are satisfied.

\begin{rema}
Evidently, these relations correspond to the usual recurrence relation and difference
equation of the $q$-Racah polynomials given, for example, in \cite{Koek}.
Let us point out that, in order to simplify the presentation and upcoming formulas,
we have moved away from the usual naming conventions of the coefficients in
the literature. The usual names can be recovered as follows:
\begin{align}
\begin{aligned}
\Phi^-(n;c_1,c_2,c_3;N)&=A_n(c_1,c_2,c_3;N)\,,\\
\Phi^+(n;c_1,c_2,c_3;N)&=C_n(c_1,c_2,c_3;N)\,,
\end{aligned}
\qquad
\begin{aligned}
\Phi^-(x;c_3,c_2,c_1;N)&=B(x;c_1,c_2,c_3;N)\,,\\
\Phi^+(x;c_3,c_2,c_1;N)&=D(x;c_1,c_2,c_3;N)\,.
\end{aligned}
\end{align}
The notation $\Phi^{\eps}$, $\eps\in\{+,0,-\}$ used in this paper makes explicit the
symmetry between the coefficients of the recurrence relation and the difference equation
and allows for more compact expressions, see \textit{e.g.}~Section \ref{sec:aw4reps}.
\end{rema}

\begin{rema}
The constraints \eqref{eq:cons1} and \eqref{eq:cons2} are necessary and sufficient
conditions to ensure $\lambda(x;c_1c_2)\neq \lambda(y;c_1c_2)$ and  $\lambda(x;c_2c_3)\neq
\lambda(y;c_2c_3)$ for $x\neq y$ and $0\leq x,y \leq N$.
\end{rema}

\subsection{Contiguity recurrence relations and difference equations \label{ssec:cont}}
The contiguity relations relate polynomials with contiguous values of $N$.
For Racah polynomials, such relations played a crucial role in constructing the
representations of the rank $2$ Racah algebra \cite{icosi}. Such contiguity relations were
also used to prove the bispectral properties of the Griffiths polynomials of Krawtchouk
type in \cite{L-griff} and of Racah type in \cite{GriffithR}. They also appear in the
study of the representations of the factorized Leonard pairs \cite{FLP}.
These relations and their proofs for the $q$-Racah polynomials are given below and will be
utilized in the study of the bispectral properties of the Tratnik polynomials of $q$-Racah
type in Section \ref{sec:Trat}.
\begin{prop} The contiguity recurrence relations between the $q$-Racah polynomials are
\begin{align}\label{eq:cont-rec-rac}
\lambda_\pm(x;c_1,c_2,c_3;N)\, \fp_n(x;c_1,c_2,c_3;N\pm 1)&= \sum_{\eps=0,\pm}
{\Phi}^{\eps}_\pm(n+\eps;c_1,c_2,c_3;N)\, \fp_{n+\eps}(x;c_1,c_2,c_3;N)\,,
\end{align}
where
\begin{subequations}\label{eq:coefrecu}
\begin{align}
\lambda_+(x;c_1,c_2,c_3;N)&=-(q^{-N-1}/c_3-q^{-x})(1-c_1c_2c_3q^{N+x+2})\,,\\
\Phi_+^+(n;c_1,c_2,c_3;N)&=-\sqrt{qc_2} c_2 q^{2n-N-1}\frac{(1-q^{n})(1-c_1q^{N-n+1})(1-c_1q^{N+2-n})(1-c_3q^{n})}{(1-c_2c_3q^{2n})(1-c_2c_3q^{2n+1})}\,,\\
\Phi_+^-(n;c_1,c_2,c_3;N)&=\, -\frac{(1-c_1c_2c_3q^{n+N+2})(1-c_1c_2c_3q^{n+N+3})(1-c_2q^{n+1})(1-c_2c_3q^{n+1})}{c_3 q^{N+1}\sqrt{qc_2}(1-c_2c_3q^{2n+1})(1-c_2c_3q^{2n+2})}\,, \\
\Phi_+^0(n;c_1,c_2,c_3;N)&=-\sum_{\eps=\pm} \Phi_+^\eps(n;c_1,c_2,c_3;N)-\frac{(1-c_1c_2q^{N+2})(1-c_1c_2c_{3}q^{N+2})}{c_3q^{N+1}\sqrt{qc_2}}\,,
\end{align}
and
\begin{align}
\lambda_-(x;c_1,c_2,c_3;N)&=-(q^{-N}-q^{-x})(1-c_1c_2q^{N+x+1})\,,\\
\Phi_-^+(n;c_1,c_2,c_3;N)&=-\sqrt{qc_2}\frac{(1-q^{n})(1-c_3q^{n})(1-c_2c_3q^{n+N})(1-c_2c_3q^{n+N+1})}{q^N(1-c_2c_3q^{2n})(1-c_2c_3q^{2n+1})}\,,\\
\Phi^-_{-}(n;c_1,c_2,c_3;N)&= -c_3\sqrt{qc_2}q^{N} \, \frac{(1-q^{n-N})(1-q^{n-N+1})(1-c_2q^{n+1})(1-c_2c_3q^{n+1})}{(1-c_2c_3q^{2n+1})(1-c_2c_3q^{2n+2})}\,,\\
\Phi^0_{-}(n;c_1,c_2,c_3;N)&=-\sum_{\eps=\pm} \Phi_-^\eps(n;c_1,c_2,c_3;N)-\sqrt{qc_2}(1-c_3q^{N})(q^{-N}-1)\,.
\end{align}
\end{subequations}
\end{prop}
\proof
The proof is done recursively on $n$. For small $n$, the relations are proven by direct
computations. Now suppose that \eqref{eq:cont-rec-rac} is valid up to a given $n$.
Compute the l.h.s.~of \eqref{eq:cont-rec-rac} for $n\mapsto n+1$, expressing
$\fp_{n+1}(x;c_1,c_2,c_3;N\pm 1)$ using the recurrence relation \eqref{eq:rec-rac}:
\begin{align}
\begin{aligned}
\lambda_\pm(x;c_1,c_2,c_3;N)\, \fp_{n+1}(x;c_1,c_2,c_3;N\pm 1)\hspace{11.5em}&\\
=\frac{\lambda_\pm(x;c_1,c_2,c_3;N)}{\Phi^+(n+1;c_1,c_2,c_3;N\pm 1)}
\Big[(\lambda(x;c_{1}c_2)
-\Phi^0(n;c_1,c_2,c_3;N\pm 1)) \, &\fp_{n}(x;c_1,c_2,c_3;N\pm 1) \\
-\Phi^-(n-1;c_1,c_2,c_3;N\pm 1) \, &\fp_{n-1}(x;c_1,c_2,c_3;N\pm 1) \Big]\,.
\end{aligned}
\end{align}
Then, use the recurrence hypothesis to get
\begin{align}
&\Phi^+(n+1;c_1,c_2,c_3;N\pm 1)\, \lambda_\pm(x;c_1,c_2,c_3;N)\,
\fp_{n+1}(x;c_1,c_2,c_3;N\pm 1)\nonumber \\
&\quad\begin{aligned}
= (\lambda(x;c_{1}c_2)-  \Phi^0(n;c_1,c_2,c_3;N\pm 1)   ) \,
&\sum_{\eps=0,\pm} {\Phi}^{\eps}_\pm(n+\eps;c_1,c_2,c_3;N)\,
\fp_{n+\eps}(x;c_1,c_2,c_3;N) \nonumber \\
-\Phi^-(n-1;c_1,c_2,c_3;N\pm 1) \,
&\sum_{\eps=0,\pm} {\Phi}^{\eps}_\pm(n-1+\eps;c_1,c_2,c_3;N)\,
\fp_{n-1+\eps}(x;c_1,c_2,c_3;N)
\end{aligned}\\
&\quad\begin{aligned}
=\sum_{\eps=0,\pm} {\Phi}^{\eps}_\pm(n+\eps;c_1,c_2,c_3;N)\,
&\sum_{\eps'=0,\pm}
{\Phi}^{\eps'}(n+\eps+\eps';c_1,c_2,c_3;N)\, \fp_{n+\eps+\eps'}(x;c_1,c_2,c_3;N)\\
-\Phi^0(n;c_1,c_2,c_3;N\pm 1)\, &\sum_{\eps=0,\pm}
{\Phi}^{\eps}_\pm(n+\eps;c_1,c_2,c_3;N)\, \fp_{n+\eps}(x;c_1,c_2,c_3;N) \\
-\Phi^-(n-1;c_1,c_2,c_3;N\pm 1)\, &\sum_{\eps=0,\pm}
{\Phi}^{\eps}_\pm(n-1+\eps;c_1,c_2,c_3;N)\, \fp_{n-1+\eps}(x;c_1,c_2,c_3;N) \,.
\end{aligned}
\end{align}
The last equation is obtained using the recurrence relation.
Finally, computing the coefficients in front of $\fp_{j}(x;c_1,c_2,c_3;N)$ for
$j=n-2,n-1,n,n+1,n+2$, relation \eqref{eq:cont-rec-rac} is proved.
\endproof
\begin{prop}The contiguity difference relations between the $q$-Racah polynomials are
\begin{align}\label{eq:cont-rec-rac2}
\lambda_\pm(n;c_3,c_2,c_1;N)\, \fp_n(x;c_1,c_2,c_3;N)&= \sum_{\eps=0,\pm}
{\Phi}^{\eps}_\pm(x;c_3,c_2,c_1;N)\, \fp_{n}(x-\eps;c_1,c_2,c_3;N\pm 1)\,,
\end{align}
where $\lambda_\pm$ and ${\Phi}^{\eps}_\pm$ are given by \eqref{eq:coefrecu}.
\end{prop}
\proof
The proof can be done following the same lines as in the previous proposition.
Another approach is to combine the previous result with the duality property.
Starting from \eqref{eq:cont-rec-rac} and replacing each $q$-Racah polynomials using
the duality relation \eqref{eq:dual} then substituting
$x\leftrightarrow n, c_1 \leftrightarrow c_3, N\mapsto N \mp 1$ and $\eps\mapsto -\eps$,
one gets
\begin{align}
\begin{aligned}
&\lambda_\pm(n;c_3,c_2,c_1;N\mp1)\,
\frac{\Omega(n;c_1,c_2,c_3;N\mp 1)}{\Omega(n;c_1,c_2,c_3;N)}\,
\fp_n(x;c_1,c_2,c_3;N)\\
&\quad=\sum_{\eps=0,\pm} \Phi_\pm^{-\eps}(x-\eps;c_3,c_2,c_1;N\mp 1)\,
\frac{\Omega(x-\eps;c_3,c_2,c_1;N\mp 1)}{\Omega(x;c_3,c_2,c_1;N)}\,
\fp_n(x-\eps;c_1,c_2,c_3;N\mp 1)\,.
\end{aligned}
\end{align}
Relation \eqref{eq:cont-rec-rac2} is proved remarking that
\begin{align}
\lambda_\pm(n;c_3,c_2,c_1;N\mp1)\,
\frac{\Omega(n;c_1,c_2,c_3;N\mp 1)}{\Omega(n;c_1,c_2,c_3;N)}
&=\alpha^\mp_N\,\, \lambda_\mp(n;c_3,c_2,c_1;N)\,,\\
\Phi_\pm^{-\eps}(x-\eps;c_3,c_2,c_1;N\mp 1)\,
\frac{\Omega(x-\eps;c_3,c_2,c_1;N\mp 1)}{\Omega(x;c_3,c_2,c_1;N)}
&=\alpha^\mp_N\,\, \Phi_\mp^{\eps}(x;c_3,c_2,c_1;N)\,,
\end{align}
where
\begin{align}\label{}
\alpha^+_N=\frac{c_1\sqrt{qc_2}(1-q^{N+1})}{1-c_1c_2c_3q^{N+2}}\,,\qquad
\alpha^-_N=\frac{1}{\alpha^+_{N-1}}\,.
\end{align}
\endproof

\section{Tratnik polynomials of $q$-Racah type}\label{sec:Trat}
A bivariate generalization of the $q$-Racah polynomials has been introduced in
\cite{GR2005,Iliev} and is here called the \textit{Tratnik polynomials of $q$-Racah type}.
We define these polynomials, recall some of their properties and provide their proofs in
the notation of this paper.

For $i,j,x,y$ nonnegative integers and $i+j\leq N$, $x+y\leq N$
the bivariate Tratnik polynomials of $q$-Racah type are defined by:
\begin{align}\label{def:Trat}
\cT_{i,j}(x,y;c_1,c_2,c_3,c_4;N)= \fp_i(x;c_1,c_2,c_3;N-j)\,\fp_j(y;c_3,c_0,c_4;N-x)\,,
\end{align}
where the parameter $c_0$ appearing in the l.h.s.~of the above relation is defined through
the following constraints between the parameters:
\begin{align}\label{eq:cont}
q^{2N+3}c_0c_1c_2c_3c_4=1\,.
\end{align}
The notation $\cT_{i,j}(x,y)$ is short for $\cT_{i,j}(x,y;c_1,c_2,c_3,c_4;N)$.
We demand that the constraints \eqref{eq:cons1}, \eqref{eq:cons3} and \eqref{eq:cons2} be
obeyed for each univariate polynomials, \textit{i.e.}~the parameters must satisfy
\begin{subequations}
\begin{gather}
c_i,q\neq 0\,,\qquad q,q^2,\dots,q^N \neq 1\,,\\
c_1c_2,\, c_2c_3,\, c_0c_3,\, c_0c_4,\, qc_1c_2c_3,\,qc_0c_3c_4 \notin
\{q^{-2},q^{-3},\dots,q^{-2N}\}\,,\\
c_0, c_1,c_2, c_3,c_4\notin \{ q^{-1},q^{-2},\dots,q^{-N} \}\,.
\end{gather}
\end{subequations}
Let us emphasize that we could multiply the previous expression of
$\cT_{i,j}(x,y;c_1,c_2,c_3,c_4;N)$ by some functions $f(i,j)$ or $g(x,y)$.
All the relations satisfied by $\cT_{i,j}(x,y;c_1,c_2,c_3,c_4;N)$ would still hold,
although with modified expressions for the coefficients occurring in the relations.

\subsection{Polynomiality}

We have been calling $\cT_{i,j}(x,y)$ ``polynomials'' without any proper
justification. We provide in the following a precise statement about the
polynomiality of $\cT_{i,j}(x,y)$.
Indeed, definition \eqref{def:Trat} of $\cT_{i,j}(x,y)$ is only valid
when $x$, $y$ are nonnegative integers and $x+y\leq N$, which defines a triangular domain in the $(x,y)$ space, we will call \textit{the grid.}
The grid is the domain over which the functions are orthogonal with respect to each other
(see Prop. \ref{pro:orthoT} below).
It is possible to multiply $\cT_{i,j}(x,y)$ by some appropriate factors of the form $f(i,j)$ and $g(x,y)$ so that
it becomes a well-defined function not only on the grid
but also for $x$, $y$ real, complex numbers or even formal variables.
More precisely, we show below that, up to transformations only valid on the grid,
this function becomes a polynomial.

Starting from definition \eqref{def:Trat} and using properties of the $q$-Pochhammer symbols, $\cT_{i,j}(x,y)$ is rewritten as follows:
\begin{equation}
\begin{split}\label{eq:tratnik2}
    \cT_{i,j}(x,y)&=\Omega(i;c_1,c_2,c_3;N-j)\,\frac{(1-c_{0}c_4q^{2j+1})}{(qc_0)^{j-\frac12(N-x)}c_3^j}\frac{(qc_0;q)_j(qc_{0}c_4;q)_j}{(q;q)_j(qc_4;q)_j}\, \frac{(qc_3;q)_{N-x}}{(qc_0c_4;q)_{N-x+1}}\\
    &\times\frac{(q^{x-N};q)_j\,\Big(\dfrac{q^{-N-x-1}}{c_{1}c_{2}};q\Big)_j}{\Big(\dfrac{q^{x-N}}{c_3};q\Big)_j\,\Big(\dfrac{q^{-N-x-1}}{c_{1}c_{2}c_{3}};q\Big)_j}
    \pphi{4}{3}{q^{-i},\; c_2c_3q^{i+1},\; q^{-x},\; c_{1}c_{2}q^{x+1}}
          {qc_2,\; c_{1}c_{2}c_{3}q^{N-j+2},\; q^{j-N}}{q;q}\\
       &\times   \pphi{4}{3}{q^{-j},\; c_{0}c_{4}q^{j+1},\; q^{-y},\; c_{0}c_{3}q^{y+1} }
          {qc_0,\; q^{N-x+2}c_3c_0c_4,\; q^{x-N}}{q;q}  \,.
\end{split}
\end{equation}
The factors in the first line of the above expression depend only on $(i,j)$ or on $x$. As explained previously, $\cT_{i,j}(x,y)$ divided by these factors satisfies the same relations as $\cT_{i,j}(x,y)$ but
with modified coefficients; therefore, we can focus on the last two lines of \eqref{eq:tratnik2}. These define a rational function w.r.t.~$\lambda(x;c_{1}c_{2})$ and $\lambda(y;c_{0}c_{3})$.
More precisely, the denominator is a polynomial of degree $j$ w.r.t.~$\lambda(x;c_{1}c_{2})$, while the numerator is a product of a polynomial of degree $i$ w.r.t.~$\lambda(x;c_{1}c_{2})$ and a bivariate polynomial of total degree $j$ w.r.t.~$\lambda(x;c_{1}c_{2})$ and $\lambda(y;c_{0}c_{3})$.

To rewrite this as a polynomial instead of a rational function, it is necessary to use two successive Sears transformations. Let us recall the formula of the Sears transformation
(see \textit{e.g.}~relation (2.10.5) in \cite{GR}):
\begin{align}
      \pphi{4}{3}{q^{-n},\;a, \;b,\;c }
          {d,\; e,\; f}{q;q}=
          \frac{(e/a;q)_n(f/a;q)_n}{(e;q)_n(f;q)_n}\;a^n
          \pphi{4}{3}{q^{-n},\;a, \;d/b,\;d/c }
          {d,\; q^{1-n}a/e,\; q^{1-n}a/f}{q;q}\,,
\end{align}
where $n$ is an integer and $abcq^{1-n}=def$.

First apply a Sears transformation on the basic hypergeometric function
${}_4\phi_3$ in the second line of \eqref{eq:tratnik2}
with the identifications $n\to x,\ a\to c_{1}c_{2}q^{x+1},\ b\to q^{-i},\ c\to c_2c_3q^{i+1},\ d\to q^{j-N},\ e\to qc_2, \ f \to c_{1}c_{2}c_{3}q^{N-j+2}$ to get
\begin{align}\label{eq:Wh1}
    \frac{(q^{-x}/c_1;q)_x(c_3q^{N-j-x+1};q)_x}{(qc_2;q)_x(c_{1}c_{2}c_{3}q^{N-j+2};q)_x}
    \pphi{4}{3}{q^{i+j-N},\;q^{j-N-i-1}c_{2}^{-1}c_{3}^{-1},\;q^{-x}, \;c_{1}c_{2}q^{x+1} }
          {qc_1,\; q^{j-N}/c_3,\; q^{j-N}}{q;q}\,.
\end{align}
Let us emphasize that this transformation is only possible
when $x$ is on the grid.

The ${}_4\phi_3$ in \eqref{eq:Wh1} is then further modified using
again the Sears transformation with the identifications $n\to N-i-j,\ a\to q^{j-N-i-1}c_{2}^{-1}c_{3}^{-1},\ b\to q^{-x},\ c\to c_{1}c_{2}q^{x+1},\ d\to q^{j-N},\ e\to qc_1,\ f\to q^{j-N}/c_3$ to get
\begin{align}
    &\frac{(c_{1}c_{2}c_{3}q^{N+2+i-j};q)_{N-i-j} (c_2q^{i+1};q)_{N-i-j} }{(qc_1;q)_{N-i-j} (q^{j-N}/c_3;q)_{N-i-j}}\nonumber\\
    \times&\pphi{4}{3}{q^{i+j-N},\;q^{j-N-i-1}c_{2}^{-1}c_{3}^{-1},\;q^{x+j-N}, \;q^{j-N-x-1}c_{1}^{-1}c_{2}^{-1} }
          {q^{2j-2N-1}c_{1}^{-1}c_{2}^{-1}c_{3}^{-1},\; q^{j-N}/c_2,\; q^{j-N}}{q;q}\,.
\end{align}
Thus, using properties of the $q$-Pochhammer symbols,
$\cT_{i,j}(x,y)$ can  be rewritten as
\begin{equation}
\begin{split}\label{eq:Tratpol}
    \cT_{i,j}(x,y)&=(-1)^{i+j} q^{\frac12i(i-1)+\frac12j(j-1)} \left(\frac{c_1c_3}{c_2c_4}\right)^{\frac12(N-x)} (qc_2)^{\frac12(N+j)} (q^{N+1}c_4)^j \\
   &\times\frac{(q^{N-i-j+1};q)_{i}}{(q;q)_{i}(q;q)_j}\,
    \frac{(1-c_{2}c_{3}q^{2i+1})}{(c_{2}c_{3}q^{i+1};q)_{N-j+1}}\,
    \frac{(qc_0;q)_j(qc_2;q)_{N-j}}{(qc_4;q)_j(c_{0}c_{4}q^{j+1};q)_j}\,
    \frac{(qc_1;q)_x}{(qc_2;q)_x}
    \\
   &\times \pphi{4}{3}{q^{i+j-N},\;q^{j-N-i-1}c_{2}^{-1}c_{3}^{-1},\;q^{x+j-N}, \;q^{j-N-x-1}c_{1}^{-1}c_{2}^{-1}}
   {q^{2j+2}c_{0}c_{4},\; q^{j-N}/c_2,\; q^{j-N}}{q;q}\\
    &\times(q^{x-N};q)_j (q^{N-x+2}c_3c_0c_4;q)_j \
          \pphi{4}{3}{q^{-j},\; c_{0}c_{4}q^{j+1},\; q^{-y},\; c_{0}c_{3}q^{y+1} }
          {qc_0,\; q^{N-x+2}c_3c_0c_4,\; q^{x-N}}{q;q}\,.
\end{split}
\end{equation}
With this expression of $\cT_{i,j}(x,y)$ one sees that, up to a renormalization by $(c_1c_3/c_2c_4)^{\frac12(N-x)}$ and $(qc_1;q)_x/(qc_2;q)_x$,
it is a polynomial of total degree $N-i$ w.r.t.~$\lambda(x;c_{1}c_{2})$ and $\lambda(y;c_{0}c_{3})$.
More precisely, it is a product of two polynomials:
a bivariate polynomial of total degree $j$ w.r.t.~$\lambda(x;c_{1}c_{2})$
and $\lambda(y;c_{0}c_{3})$, multiplied by
a polynomial of degree $N-i-j$ w.r.t.~$\lambda(x;c_{1}c_{2})$.

\begin{rema}
The Tratnik polynomials of $q$-Racah type \eqref{def:Trat} are in correspondence with the
definition (2.13) of \cite{Iliev}. The expression $R_2$ of the Tratnik polynomials in
\cite{Iliev} is given by, for $n_1$, $n_2$, $z_1$ and $z_2$ nonnegative integers, $n_1+n_2\leq N$ and $0\leq z_1\leq z_2\leq N$,
\begin{equation}
\begin{split}
& R_2(n_1,n_2;z_1,z_2;\vec{\alpha};N) = \frac{(\frac{\alpha_1^2}{\alpha_0^2}, \alpha_2^2q^{z_2}, q^{-z_2};q)_{n_1}}
{\alpha_1^{n_1}} \,\pphi{4}{3}{q^{-n_1},\, q^{n_1}\frac{\alpha_2^2}{q\alpha_0^2},\; q^{-z_1},\; \alpha_1^2q^{z_1}}
                             {\frac{\alpha_1^2}{\alpha_0^2},\; \alpha_2^2q^{z_2},\; q^{-z_2}}{q;q} \\
&\times\, \frac{(q^{2n_1}\frac{\alpha_2^2}{\alpha_0^2},\; q^{N+n_1}\alpha_3^2,\; q^{n_1-N};q)_{n_2} }{(q^{n_1}\alpha_2)^{n_2}}
\,\pphi{4}{3}{q^{-n_2},\; q^{2n_1+n_2}\frac{\alpha_3^2}{q\alpha_0^2},\; q^{n_1-z_2},\;q^{n_1+z_2}\alpha_2^2}
             {q^{2n_1}\frac{\alpha_2^2}{\alpha_0^2},\; q^{N+n_1}\alpha_3^2,\; q^{n_1-N}}{q;q}\,,
\end{split}
\end{equation}
where $\vec{\alpha} = (\alpha_0,\alpha_1,\alpha_2,\alpha_3)$.

Let us relate $n_1,n_2,z_1,z_2$ and the parameters $\alpha_i$
to $i,j,x,y$ and the parameters $c_i$ through
\begin{gather}
\begin{gathered}
z_1 = y\,,\qquad z_2 = N-x\,,\qquad n_1 = j\,,\qquad n_2 = N-i-j\,,\\
c_1=\frac{\alpha_3^2}{q\alpha_2^2} \,, \qquad
c_2=\frac{1}{q^{2N}\alpha_3^2} \,, \qquad
c_3 = \alpha_0^2 \,, \qquad
c_4=\frac{\alpha_2^2}{q\alpha_1^2} \,, \qquad
c_0=\frac{\alpha_1^2}{q\alpha_0^2} \,.
\end{gathered}
\end{gather}
Then, the Tratnik polynomials of $q$-Racah type defined in the above equation \eqref{eq:Tratpol}
are linked to $R_2$ as follows:
\begin{align}\label{eq:TratIlv}
    & R_2(n_1,n_2;z_1,z_2;\vec{\alpha};N)=(-1)^{i+j}\left(\frac{c_1c_3}{c_2c_4}\right)^{\frac12(x-N)}
    \left(q^{2i+1}\frac{c_2}{c_1}\right)^{\frac12(i-N)}
    \left(\frac{c_2}{c_4}\right)^{\frac12(j-N)} (qc_4)^{-N/2}
    \\
   &\qquad \times \frac{(q;q)_j(q;q)_{N-i-j}}{q^{\frac12i(i-1)+\frac12j(j-1)-ij}} \,
    \frac{(c_{2}c_{3}q^{i+1};q)_{N-j+1}}{(1-c_{2}c_{3}q^{2i+1})(qc_2;q)_{i}} \,
    \frac{(c_{0}c_{4}q^{j+1};q)_{N-i+1}(qc_4;q)_j}{(1-c_{0}c_{4}q^{2j+1})} \,
    \frac{(qc_2;q)_x}{(qc_1;q)_x} \; \cT_{i,j}(x,y)\,.
    \nonumber
\end{align}
\end{rema}

\subsection{Orthogonality and duality}

The following identities are used to prove several properties of the Tratnik polynomials:
\begin{lemm}
The function $\Lambda$
\begin{align}
\label{eq:Lambda}
\Lambda(x;c_1,c_2,c_3,c_4;N) &=\frac{(q^{-N};q)_{x}}{(q;q)_{x}}
\frac{(1-c_1c_2q^{2x+1})(qc_2;q)_x (q^{2x+1}c_2)^{N/2}}{(qc_1;q)_x(q^{x+1}c_1c_2;q)_{N+1}}
\left( \frac{c_1c_3}{c_2c_4}  \right)^{\!x/2} \,,
\end{align}
as well as the functions $\Omega$ and $\omega$ defined respectively in \eqref{def:Omega} and \eqref{eq:omega} obey the following relations:
\begin{align}
\frac{\Omega(x;c_3,c_2,c_1;N-j)}{\Omega(j;c_3,c_0,c_4;N-x)}
&= \frac{\Lambda(x;c_1,c_2,c_3,c_4;N)}{\Lambda(j;c_4,c_0,c_3,c_1;N)}\,,
\label{eq:*}\\
\frac{\Lambda(x;c_1,c_2,c_3,c_4;N)}{\Lambda(x;c_1,c_2,c_4,c_3;N)}
&= \frac{\omega_x\Big(\frac{c_1c_2c_3}{c_4}\Big)}{\omega_x\Big(\frac{c_1c_2c_4}{c_3}\Big)}
\label{eq:**}\,.
\end{align}
\end{lemm}
\proof
Straightforward calculation from the explicit expressions \eqref{def:Omega},
\eqref{eq:omega} and \eqref{eq:Lambda}.
\endproof

\begin{prop}[\textbf{Orthogonality}]\label{pro:orthoT}
The Tratnik polynomials obey the following orthogonality relation, for $i,j,k,\ell$
nonnegative integers such that $i+j\leq N$ and $k+\ell\leq N$:
\begin{align}
&\sum_{\atopn{0\leq x,y\leq N}{x+y\leq N}} \Omega(y;c_4,c_0,c_3;N-x)\,
\Lambda(x;c_1,c_2,c_3,c_4;N)\,\cT_{i,j}(x,y;c_1,c_2,c_3,c_4;N)\,
\cT_{k,\ell}(x,y;c_1,c_2,c_3,c_4;N)\nonumber\\
&\qquad\qquad=
\delta_{i,k}\,\delta_{j,\ell}\,\,
\Omega(i;c_1,c_2,c_3;N-j)\,\Lambda(j;c_4,c_0,c_3,c_1;N)\,.
\end{align}
\end{prop}
\proof
Direct calculation, using the explicit expression \eqref{def:Trat} for the Tratnik
polynomials, the relation \eqref{eq:*} and the orthogonality relation
\eqref{eq:ortho} for the Racah polynomials.
\endproof
As for the univariate polynomials, the previous relation defines orthogonal polynomials if
the parameters are chosen such that the weight is strictly positive.

\begin{prop}[\textbf{Duality}]
The Tratnik polynomials obey the following duality relation, for $i,j,x,y$
nonnegative integers such that $i+j\leq N$ and $x+y\leq N$:
\begin{equation}\label{eq:dualT}
\frac{\cT_{i,j}(x,y;c_1,c_2,c_3,c_4;N)}
     {\Omega(i;c_1,c_2,c_3;N-j)\,\Lambda(j;c_4,c_0,c_3,c_1;N)}=
\frac{\cT_{y,x}(j,i;c_4,c_0,c_3,c_1;N)}
     {\Omega(y;c_4,c_0,c_3;N-x)\,\Lambda(x;c_1,c_2,c_3,c_4;N)}\,.
\end{equation}
\end{prop}
\proof
Consequence of relation \eqref{eq:*} combined with the duality relation \eqref{eq:dual}
for the Racah polynomials.
\endproof
As for the univariate Racah polynomials, we can compose the orthogonality and duality
relations to obtain an involution relation for the Tratnik polynomials:
\begin{equation}\label{eq:involT}
\sum_{\atopn{0\leq x,y\leq N}{x+y\leq N}}\cT_{i,j}(x,y;c_1,c_2,c_3,c_4;N)\,
\cT_{y,x}(\ell,k;c_4,c_0,c_3,c_1;N) =
\delta_{i,k}\,\delta_{j,\ell}\,.
\end{equation}
It has been established in \cite{Iliev} that these bivariate polynomials are bispectral.
The bispectral properties are recalled in the following propositions along with, to the
best of our knowledge, new proofs.

To simplify the notations, we define the following two functions:
\begin{align}
F^-_x(c_1,c_2)&=c_1c_2q^{x+1}
\frac{(1-c_2q^{x+1})(1-c_1c_2q^{x+1})}{(1-c_1c_2q^{2x+1})(1-c_1c_2q^{2x+2})}\,,\\
F^+_x(c_1,c_2)&=c_1c_2^2q^{x+1}
\frac{(1-c_1q^{x})(1-q^{x})}{(1-c_1c_2q^{2x})(1-c_1c_2q^{2x+1})}\,.
\end{align}

\subsection{Bispectral properties}

\begin{prop}[\textbf{Recurrence relations}]\label{prop:rec-trat}
For $i,j,x,y$ nonnegative integers such that $x+y\leq N$ and $i+j\leq N$,
the Tratnik polynomials obey the following recurrence relations:
\begin{align}
\lambda(x;c_{1}c_2)\,\cT_{i,j}(x,y) &= \sum_{\eps=0,\pm}
\Phi^\eps(i+\eps;c_1,c_2,c_3;N-j)\,\cT_{i+\eps,j}(x,y) \,,
\label{eq:rec1-Trat}\\
\lambda(y;c_{3}c_0)\,\cT_{i,j}(x,y) &= \sum_{\eps,\eps'=0,\pm}
A^{\eps,\eps'}_{i+\eps,j+\eps'}(c_1,c_2,c_3,c_4;N)\,\cT_{i+\eps,j+\eps'}(x,y)\,,
\label{eq:rec2-trat}
\end{align}
where $\lambda$ and $\Phi^\eps$ are given by \eqref{eq:rec-rac-coeff} and
\begin{subequations}
\label{def:Aij}
\begin{align}
A_{i,j}^{\eps,\eps'}(c_1,c_2,c_3,c_4;N) &=
-c_3q^{N+1}\,F^{\eps'}_j(c_4,c_0)\,
\Phi_{\eps'}^\eps(i;c_1,c_2,c_3;N-j)\,,
\qquad\text{for $\eps=0,\pm$, $\eps'=\pm$\,,}\\
A_{i,j}^{\eps,0}(c_1,c_2,c_3,c_4;N) &=
c_3q^{N+1}\Big(\!F^{+}_j(c_4,c_0)+F^{-}_j(c_4,c_0)\!\Big)\,
\Phi^\eps(i;c_1,c_2,c_3;N-j)\,,
\quad\text{\,for $\eps=\pm$\,,}\\
A_{i,j}^{0,0}(c_1,c_2,c_3,c_4;N) &=
-\frac{(1-c_3q^{N+1-j})\sqrt{qc_2} }{1-c_1 c_2q^{N+2-j}}
\sum_{\eps=0,\pm} A_{i,j}^{\eps,+}(c_1,c_2,c_3,c_4;N) \label{eq:A00rel}\\
&\hspace{1em}-\frac{1-c_1c_2q^{N+1-j} }{(1-c_3q^{N-j})\sqrt{qc_2}}
\sum_{\eps=0,\pm} A_{i,j}^{\eps,-}(c_1,c_2,c_3,c_4;N)
-\sum_{\eps=\pm} A_{i,j}^{\eps,0}(c_1,c_2,c_3,c_4;N)\,.\nonumber
\end{align}
\end{subequations}
We apply the convention $\cT_{-1,j}(x,y)=\cT_{i,-1}(x,y)=\cT_{-1,-1}(x,y)=0$.
\end{prop}
\proof
The equation \eqref{eq:rec1-Trat} is a direct consequence of the recurrence equation
\eqref{eq:rec-rac} applied to the Racah polynomial $\fp_i(x;c_1,c_2,c_3;N-j)$ in
$\cT_{i,j}(x,y)$.

To prove \eqref{eq:rec2-trat}, apply the recurrence equation to the second polynomial
$\fp_j(y;c_3,c_0,c_4;N-x)$ such that the l.h.s.~of \eqref{eq:rec2-trat} becomes
\begin{align}\label{eq:pr1}
&\lambda(y;c_{3}c_0)\,\cT_{i,j}(x,y) = \fp_i(x;c_1,c_2,c_3;N-j)
\sum_{\eps=0,\pm} \Phi^\eps(j+\eps;c_3,c_0,c_4;N-x)\,\fp_{j+\eps}(y;c_3,c_0,c_4;N-x)\,.
\end{align}
Remark that for $\eps=\pm$,
\begin{align}
\Phi^\eps(j+\eps;c_3,c_0,c_4;N-x) &= -c_3q^{N+1}F^\eps_{j+\eps}(c_4,c_0)\,
\lambda_\eps(x;c_1,c_2,c_3;{N-j-\eps})\,,
\label{eq:Ph0gj}\\
\Phi^0(j;c_3,c_0,c_4;N-x) &= c_3q^{N+1}\Big(F_j^+(c_4,c_0)+F_j^-(c_4,c_0)\Big)
\lambda(x;c_1c_2) + g_j\,,
\end{align}
where
\begin{align}
\label{eq:gj}
g_j = - (1-c_3q^{N+1-j}) (q^j-c_1c_2c_3q^{N+2}) F^+_j(c_4,c_0)
      - c_3 (1-q^{N-j}) (q^{j+1}-c_1c_2q^{N+2}) F^-_j(c_4,c_0) \,.
\end{align}
Then, apply the recurrence and contiguity recurrence relations \eqref{eq:rec-rac} and
\eqref{eq:cont-rec-rac} to the first polynomial $\fp_i(x;c_1,c_2,c_3;N-j)$ in
\eqref{eq:pr1}.
This allows to recognize Tratnik polynomials with shifted indices as written in
\eqref{eq:rec2-trat}, with the values given in \eqref{def:Aij}, noticing that
\begin{align}
\label{eq:gjAij}
g_j = - \frac{ (1-c_3q^{N+1-j}) \sqrt{qc_2} }{ 1-c_1 c_2q^{N+2-j} }
        \sum_{\eps=0,\pm} A_{i,j}^{\eps,+}
      - \frac{ 1-c_1c_2q^{N+1-j} }{ (1-c_3q^{N-j}) \sqrt{qc_2} }
        \sum_{\eps=0,\pm} A_{i,j}^{\eps,-} \,,
\end{align}
where the notation $A_{i,j}^{\eps,\pm}$ is short for
$A_{i,j}^{\eps,\pm}(c_1,c_2,c_3,c_4;N)$.
\endproof
Note that $\cT_{N+1-j,j}(x,y)=0$ due to the properties of the $q$-Racah polynomials \eqref{eq:fctpn}.
\begin{rema}
Note that the normalization of the Tratnik polynomials can be changed in such a way that
the relation \eqref{eq:A00rel} simplifies to
\begin{align}\label{eq:stock}
\sum_{\eps,\eps'=0,\pm} \widetilde A_{i,j}^{\eps,\eps'}(c_1,c_2,c_3,c_4;N) =0\,.
\end{align}
Indeed, by changing the normalization of the Tratnik polynomials as
\begin{align}
\cT_{i,j}(x,y) \mapsto \frac{(c_1c_2q^{N-j+2};q)_j }{(qc_2)^{j/2} (c_3q^{N-j+1};q)_j}
\cT_{i,j}(x,y)
\end{align}
the coefficients of the first recurrence relation \label{eq:rec1-trat} for this
renormalized Tratnik polynomial are unchanged whereas the ones of
the second recurrence relation \eqref{eq:rec2-trat} are modified as follows:
\begin{align}
A_{i,j}^{\eps,\eps'}(c_1,c_2,c_3,c_4;N) \mapsto
\widetilde A_{i,j}^{\eps,\eps'}(c_1,c_2,c_3,c_4;N)
 = \left(\!\frac{\sqrt{qc_2}(1-c_3q^{N-j+\frac12+\frac{\eps'}{2}})}
                {1-c_1c_2q^{N-j+\frac32+\frac{\eps'}{2}}}\!\right)^{\eps'}\!\!
A_{i,j}^{\eps,\eps'}(c_1,c_2,c_3,c_4;N)\,.
\end{align}
The relation \eqref{eq:stock} opens the path to connection to stochastic models, such as
birth and death processes, in the spirit of \textit{e.g.}~\cite{Sasaki2022}.
\end{rema}
In addition of the two recurrence relations, the Tratnik polynomials satisfy two
difference equations given below.
\begin{prop}[\textbf{Difference equations}]
For $x,y,i,j$ nonnegative integers such that $x+y\leq N$ and $i+j\leq N$,
the Tratnik polynomials satisfy the following two difference equations:
\begin{align}\label{eq:DiffT1}
\lambda(j;c_4c_{0})\,\cT_{i,j}(x,y) &= \sum_{\eps=0,\pm}
\Phi^\eps(y;c_4,c_0,c_3;N-x)\,\cT_{i,j}(x,y-\eps) \,,\\
\label{def:Dij}
\lambda(i;c_{2}c_3)\,\cT_{i,j}(x,y) &= \sum_{\eps,\eps'=0,\pm}
A^{\eps,\eps'}_{y,x}(c_4,c_0,c_3,c_1;N)\,\cT_{i,j}(x-\eps',y-\eps)\,.
\end{align}
\end{prop}
\proof
The equation \eqref{eq:DiffT1} is a direct consequence of the difference relation
\eqref{eq:rec-rac4} applied to the Racah polynomial $\fp_j(y;c_3,c_0,c_4;N-x)$ in
$\cT_{i,j}(x,y)$.
For relation \eqref{def:Dij}, the calculation may proceed along the same lines as what was
done for Proposition \ref{prop:rec-trat}. However, it can also be proven by combining the
previous result \eqref{eq:rec2-trat}, the duality relation \eqref{eq:dualT} and the
following relations between the coefficients:
\begin{align}
\begin{aligned}
A_{y,x}^{\eps,\eps'}(c_4,c_0,c_3,c_1;N) &=
\frac{\Omega(y-\eps;c_4,c_0,c_3;N-x+\eps')\Lambda(x-\eps';c_1,c_2,c_3,c_4;N)}
     {\Omega(y;c_4,c_0,c_3;N-x)\Lambda(x;c_1,c_2,c_3,c_4;N)}\\
&\times A_{y-\eps, x-\eps'}^{-\eps,-\eps'}(c_4,c_0,c_3,c_1;N)\,,
\end{aligned}
\end{align}
where $\eps,\eps'=0,\pm$.
\endproof

\begin{rema}[\textbf{Generalized Favard theorem}]\label{rem:Favard-T}
A multivariate version of the Favard theorem has been established, see
\textit{e.g.}~Theorem 3.2.8 in \cite{DuXu}. It provides necessary and sufficient
conditions for a family of multivariate polynomials satisfying recurrence relations to be
orthogonal. For bivariate polynomials $P_{i,j}(x_1,x_2)$ of total degree $i+j$, the
recurrence relations must be of the following form, for $i=1,2$:
\begin{equation}\label{eq:rec-Favard}
x_i\,\PP_{n} = \AA_{n,i}\,\PP_{n+1}+\BB_{n,i}\,\PP_{n}+\AA^ T_{n-1,i}\,\PP_{n-1}\,.
\end{equation}
In the above, we gather the $r_n$ polynomials of total degree $n$ in a vector $\PP_{n}$.
In the r.h.s., $\AA_{n,i}$ and $\BB_{n,i}$ are matrices of size
$r_n\times r_{n+1}$ and $r_n\times r_{n}$ respectively.
An additional necessary condition is that $\BB_{n,i}$ must be symmetric.

We now show that these conditions are satisfied for the bivariate Tratnik polynomials.
For nonnegative integers $i,j$ such that $i+j\leq N$,
one defines
\begin{align}\label{eq:Tbar}
\overline\cT_{i,j}(x,y)&=\cT_{i,j}(x,y)\,/\,\cT_{N,0}(x,y)
\end{align}
where $\cT_{i,j}(x,y)$ is given by \eqref{eq:Tratpol}.
It is of total degree $N-i$ w.r.t.~$\lambda(x;c_1c_2)$ and $\lambda(y;c_0c_3)$.
Consider the vectors
\begin{equation}
\TT_{n}=\Big(\nu_{n,0}\overline\cT_{N-n,0}\,,\ \nu_{n,1}\overline\cT_{N-n,1}\,,\ \dots\,,\
\nu_{n,n}\overline\cT_{N-n,n}\Big)^T
\end{equation}
containing all polynomials of total degree $n$ renormalized by some coefficients $\nu_{i,j}$.
In order to get the recurrence relations \eqref{eq:rec2-trat} and \eqref{eq:rec1-Trat}
in the form \eqref{eq:rec-Favard}, we need to impose
\begin{gather}\label{eq:cont-nu}
\begin{gathered}
\left(\frac{\nu_{i,j+1}}{\nu_{i,j}}\right)^2
= \frac{A^{0,+}_{N-i,j+1}(c_1,c_2,c_3,c_4;N)}{A^{0,-}_{N-i,j}(c_1,c_2,c_3,c_4;N)}\,,
\qquad \left(\frac{\nu_{i+1,j}}{\nu_{i,j}}\right)^2
= \frac{A^{-,0}_{N-i-1,j}(c_1,c_2,c_3,c_4;N)}{A^{+,0}_{N-i,j}(c_1,c_2,c_3,c_4;N)}\,,\\
\left(\frac{\nu_{i+1,j+1}}{\nu_{i,j}}\right)^2
= \frac{A^{-,+}_{N-i-1,j+1}(c_1,c_2,c_3,c_4;N)}{A^{+,-}_{N-i,j}(c_1,c_2,c_3,c_4;N)}\,,
\qquad \left(\frac{\nu_{i,j+1}}{\nu_{i+1,j}}\right)^2
= \frac{A^{+,+}_{N-i,j+1}(c_1,c_2,c_3,c_4;N)}{A^{-,-}_{N-i-1,j}(c_1,c_2,c_3,c_4;N)}\,,\\
\left(\frac{\nu_{i+1,j}}{\nu_{i,j}}\right)^2
= \frac{\Phi^-(N-i-1;c_1,c_2,c_3;N-j)}{\Phi^+(N-i;c_1,c_2,c_3;N-j)}\,,
\end{gathered}
\end{gather}
where $A_{i,j}^{\eps,\eps'}(c_1,c_2,c_3,c_4;N)$ and $\Phi^\eps(i+\eps;c_1,c_2,c_3;N-j)$
are the coefficients introduced in \eqref{def:Aij} and \eqref{eq:rec-rac-coeff}.
Using the explicit expression of these coefficients, one can show that in
\eqref{eq:cont-nu}, the second and third lines are a consequence of the first line, the
latter being sufficient to determine the coefficients $\nu$ up to a global constant.
Thus, one concludes from the generalized Favard theorem that the polynomials $\overline\cT_{i,j}(x,y)$ are orthogonal, in accordance with the
result in Proposition \ref{pro:orthoT}.
\end{rema}

\section{Racah and Biedenharn--Elliot relations for $q$-Racah polynomials}
\label{sec:RBE}
Two fundamental relations involving univariate $q$-Racah polynomials are the Racah and
Biedenharn--Elliot relations. These relations will be key ingredients in
the next section.
Both these relations have been obtained in the study of $6j$-symbols of the quantum
algebra $U_q(su(2))$, meaning that they were valid only for a restricted range of
parameters.
We present them below in their full generality.
To the best of our knowledge, these general relations have not appeared in the literature
before and the proofs we provide are new.

\subsection{Racah relation}\label{ssec:Rrelation}
The Racah relation appeared in the study of the $6j$-symbol for the Lie algebra $su(2)$
(see \textit{e.g.}~\cite{Mess}) and has been related to the Racah algebra
(a limit $\scq\to 1$ of the Askey--Wilson algebra) in \cite{Zh2}.
As the $6j$-symbol can be expressed in terms of Racah polynomials \cite{Wilson, KMV},
the Racah relation leads to an identity for Racah polynomials with particular parameters.
This relation was then generalized through the study of the representations of
the Racah algebra \cite{icosi}.
When the $6j$-symbol for the quantum group $U_q(su(2))$ is instead considered,
there exists a similar relation involving the $q$-Racah polynomials with
particular parameters \cite{KR} and their study is intimately linked to
the Askey--Wilson algebra $\aw$ \cite{GZ}.
In this paper, using the representations of the Askey--Wilson algebra $\aw$ introduced
in the Appendix \ref{app:aw3} and following the lines of the proof in the non-deformed
case \cite{icosi}, we prove the Racah relation for generic $q$-Racah polynomials.
\begin{prop} The Racah relation between the $q$-Racah polynomials reads,
for $0\leq i,x \leq N$,
\begin{align}\label{eq:racahrel}
\begin{aligned}
\sum_{n=0}^N\ &
\frac{\omega_n(c_1c_2)}{\xi_n(c_1,c_2)}
\frac{\omega_i(c_2c_3)}{\xi_i(c_2,c_3)}
\frac{\omega_x(c_3c_1)}{\xi_x(c_3,c_1)}\,
\fp_i(n;c_1,c_2,c_3;N)\,
\fp_n(x;c_3,c_1,c_2;N)\\
&=\omega_N(qc_1c_2c_3)\,
\fp_i(x;c_1,c_3,c_2;N)\,,
\end{aligned}
\end{align}
where the functions $\xi$ and $\omega$ are defined in \eqref{eq:omega}.
\end{prop}
\proof
Consider the three equivalent representations of $\aw$ $\cA$, $\cB$, $\cC$
introduced in the Appendix \ref{sec:aw3reps}.
The changes of basis between these representations \eqref{eq:aw3_isoreps} lead to
\begin{align}
\cP_{\!\!\scriptscriptstyle \cC\cB}\,\cB_{I}\,\cP_{\!\!\scriptscriptstyle \cB\cC}
= \cC_{I} =
\cP_{\!\!\scriptscriptstyle \cC\cA}\, \cP_{\!\!\scriptscriptstyle \cA\cB}\,
\cB_{I}\,
\cP_{\!\!\scriptscriptstyle \cB\cA}\, \cP_{\!\!\scriptscriptstyle \cA\cC}\,,
\qquad I\in\{12,23,13\}\,.
\end{align}
Since the representations $\cA$, $\cB$ and $\cC$ are irreducible,
the above formula implies that there exists (by Schur's Lemma) a constant $\tau$
such that
\begin{align}\label{eq:ppp}
\cP_{\!\!\scriptscriptstyle \cB\cA}\, \cP_{\!\!\scriptscriptstyle \cA\cC}
=\tau\, \cP_{\!\!\scriptscriptstyle \cB\cC}\,.
\end{align}
To determine the constant $\tau$ take the component $(0,0)$ of the previous equation.
This leads to:
\begin{align}
\tau = \left(\frac{qc_1c_2}{c_3}\right)^{\!\frac{N}{2}}
\sum_{k=0}^N q^{Nk}\,\frac{(q^{-N};q)_{k}}{(q;q)_{k}}
\frac{(1-c_1c_2q^{2k+1})(c_1c_2c_3q^{N+2};q)_k(qc_3;q)_{N-k}}
     {(c_1c_2q^{k+1};q)_{N+1}}\,.
\end{align}
The above sum can be computed from the orthogonality relation of the $q$-Racah
polynomials. Starting from \eqref{eq:ortho} in the case where $n=m=0$,
replace $c_1\mapsto c_1/t$ and $c_2\mapsto c_2 t$ and take the limit $t\to\infty$.
This allows to identify
\begin{align}
 \tau=\omega_N(qc_1c_2c_3)\,.
\end{align}
Finally, the entry $(i,x)$ of relation \eqref{eq:ppp} is
precisely the Racah relation of the proposition.
\endproof

\subsection{Biedenharn--Elliot relation}
\label{ssec:BErelation}
Similarly to the Racah relation, the Biedenharn--Elliot relation is a relation between the
$6j$-symbols for the Lie algebra $su(2)$ \cite{Bie,Ell} which has been generalized to the
case of
the quantum group $U_q(su(2))$ in \cite{KR,MMS}. In \cite{icosi}, such a relation is also
obtained in the context of the Racah algebra. We provide here this relation for generic
$q$-Racah polynomials.
\begin{prop}
For $i,j,x,y$ nonnegative integers such that $0\leq i+j\leq N$ and $0\leq x+y\leq N$,
the Biedenharn--Elliot relation between the $q$-Racah polynomials is:
\begin{align}\label{eq:biedenrel}
\begin{aligned}
\sum_{a=0}^{\min(N-j,N-y)} &\;
\fp_{i}(a;c_3,c_2,c_1;N-j)\,
\fp_{j}(y;c_1,c_0,c_4;N-a)\,
\fp_{a}(x;c_4,c_3,c_2;N-y)
\\[-.75em]
& \times
\big(\tfrac{c_0}{c_3}\big)^{\!\frac i2}
\big(\tfrac{c_3}{c_1}\big)^{\!\frac j2}
\big(\tfrac{c_1}{c_4}\big)^{\!\frac a2}
\big(\tfrac{c_4}{c_2}\big)^{\!\frac y2}
\big(\tfrac{c_2}{c_0}\big)^{\!\frac x2}\,
\xi_{i}(c_1,c_2)\,
\xi_{j}(c_4,c_0)\,
\xi_{a}(c_2,c_3)\,
\xi_{y}(c_0,c_1)\,
\xi_{x}(c_3,c_4)
\\[1em]
& =
\fp_{j}(x;c_3,c_4,c_0;N-i)\,\fp_{i}(y;c_0,c_1,c_2;N-x)
\,.
\end{aligned}
\end{align}
The last line of the equation can also be written as a Tratnik polynomial
$\cT_{j,i}(x,y;c_3,c_4,c_0,c_2;N)$.
\end{prop}
\proof
Consider the five equivalent representations of $\haw$ $\sA$, $\sB$, $\sC$, $\sD$, $\sE$
introduced in the Appendix \ref{sec:aw4reps}.
The changes of basis between these representations \eqref{eq:aw4_isoreps} lead to
\begin{align}
\sP_{\!\!\scriptscriptstyle \sA\sB}\,
\sP_{\!\!\scriptscriptstyle \sB\sC}\,
\sP_{\!\!\scriptscriptstyle \sC\sD}\,
\sD_{I}\,
\sP_{\!\!\scriptscriptstyle \sD\sC}\,
\sP_{\!\!\scriptscriptstyle \sC\sB}\,
\sP_{\!\!\scriptscriptstyle \sB\sA}\,
= \sA_{I} =
\sP_{\!\!\scriptscriptstyle \sA\sE}\,
\sP_{\!\!\scriptscriptstyle \sE\sD}\,
\sD_{I}\,
\sP_{\!\!\scriptscriptstyle \sD\sE}\,
\sP_{\!\!\scriptscriptstyle \sE\sA}\,,
\qquad I\in\{12,23,34,123,234\}\,.
\end{align}
Since the representations $\sA$, $\sB$, $\sC$, $\sD$ and $\sE$ are irreducible,
the above formula implies that there exists (by Schur's Lemma) a constant $\tau$
such that
\begin{align}\label{eq:ppppp}
\sP_{\!\!\scriptscriptstyle \sA\sB}\,
\sP_{\!\!\scriptscriptstyle \sB\sC}\,
\sP_{\!\!\scriptscriptstyle \sC\sD}\,
=\tau\,
\sP_{\!\!\scriptscriptstyle \sA\sE}\,
\sP_{\!\!\scriptscriptstyle \sE\sD}\,.
\end{align}
To determine the constant $\tau$ take the component $(0,0);(0,0)$ of
the previous equation. This leads to:
\begin{align}
\tau = \frac{\Omega(0;c_3,c_2,c_1;N)}{\Omega(0;c_3,c_4,c_0;N)\,\Omega(0;c_0,c_1,c_2;N)}
\sum_{j=0}^N \Big(\frac{c_1}{c_4}\Big)^{\!\frac{j}{2}}
\xi_j(c_2,c_3)\,\Omega(0;c_1,c_0,c_4;N-j)\,\Omega(j;c_4,c_3,c_2;N)\,.
\end{align}
The above sum can be computed from the orthogonality relation of the $q$-Racah
polynomials. Starting from \eqref{eq:ortho} in the case where $n=m=0$,
replace $c_1\mapsto 1/(c_1q^{N+1})$, $c_2\mapsto c_1c_2c_3q^{N+1}$, $c_3\mapsto c_4$.
We find that $\tau$ is precisely equal to
\begin{align}
\tau=1\,.
\end{align}
Finally, the entry $(i,j);(y,x)$ of relation \eqref{eq:ppppp} is
precisely the Biedenharn--Elliot relation of the proposition.
\endproof

\section{Griffiths polynomials of $q$-Racah type}\label{sec:Griff}
We now introduce the main functions of this paper, called the
\textit{Griffiths polynomials of $q$-Racah type}.
For $i,j,x,y$ nonnegative integers such that $i+j,x+y\leq N$, they are defined as:
\begin{align}
&\cG_{i,j}(x,y;c_1,c_2,c_3,c_4;N)\nonumber\\
&\qquad =\sum_{a=0}^{M} \omega_a\Big(\frac{c_1c_2c_3}{c_4}\Big)\,
\fp_i(a;c_1,c_2,c_3;N-j)\,\fp_j(y;c_3,c_0,c_4;N-a)\,\fp_a(x;c_4,c_2,c_1;N-y) \,,
\label{eq:G3p}
\end{align}
where $M=\min(N-j,N-y)$. Due to the properties of the polynomials $p_n(x;c_1,c_2,c_3;N)$
(see beginning of Section \ref{sec:Rac}), the terms with $a > \min(N-j,N-y)$ \textit{do not} contribute to the sum in \eqref{eq:G3p}, and one can replace $M$ by either $N-j$ or $N-y$.
Again, the parameter $c_0$ is fixed through the constraint
equation $c_0c_1c_2c_3c_4q^{2N+3}=1$ and $\omega_a(c)$ is given by \eqref{eq:omega}.

The Griffiths polynomials of $q$-Racah type can be expressed in terms of the
Tratnik polynomials introduced previously:
\begin{align}
\cG_{i,j}(x,y)
&=\sum_{a=0}^{N-y} \omega_a\Big(\frac{c_1c_2c_3}{c_4}\Big)\,
\cT_{i,j}(a,y;c_1,c_2,c_3,c_4;N)\,\fp_a(x;c_4,c_2,c_1;N-y)\label{eq:GTp}\\
& = \sum_{a=0}^{N-j} \omega_a\Big(\frac{c_1c_2c_3}{c_4}\Big)\, \fp_i(a;c_1,c_2,c_3;N-j)\,
\cT_{j,a}(y,x;c_3,c_0,c_4,c_1;N)\,.\label{eq:GpT}
\end{align}
The notation $\cG_{i,j}(x,y)$ is short for the functions
$\cG_{i,j}(x,y;c_1,c_2,c_3,c_4;N)$. \\
We demand that, for each univariate polynomials in the definition of the Griffiths
polynomials \eqref{eq:G3p}, the constraints \eqref{eq:cons1}, \eqref{eq:cons3} and
\eqref{eq:cons2} be obeyed, \textit{i.e.} the parameters must satisfy
\begin{subequations}\label{eq:cond-G}
\begin{gather}
c_i,q\neq 0\,,i=0,1,\dots,4\,,\qquad q,q^2,\dots,q^N \neq 1\,,\\
c_1c_2,\, c_2c_3,\, c_0c_3,\, c_0c_4,\, c_2c_4,\, qc_1c_2c_3,\, qc_0c_3c_4,\, qc_1c_2c_4
\notin \{q^{-2},q^{-3},\dots,q^{-2N}\}\,,\\
c_0,c_1,c_2,c_3,c_4\notin \{ q^{-1},q^{-2},\dots,q^{-N} \}\,.
\end{gather}
\end{subequations}

Replacing $c_i\to q^{c_i}$ and taking the limit $q\to 1$ in \eqref{eq:G3p}, we recover the
\textit{Griffiths polynomial of Racah type} studied in \cite{GriffithR}:
\begin{equation}
\lim_{q\to1}\, \cG_{i,j}(x,y;q^{c_1},q^{c_2},q^{c_3},q^{c_4};N)
=G_{i,j}(x,y;c_1,c_2,c_3,c_4;N)\,.
\end{equation}
The polynomials $G_{i,j}(x,y)$ were introduced in \cite{icosi} with a different
normalization. As mentioned in the introduction, it has been shown in \cite{GriffithR}
that the polynomials $G_{i,j}(x,y;c_1,c_2,c_3,c_4;N)$ tend (in a certain limit) to the
original bivariate Griffiths polynomials introduced in \cite{Griff}. This latter fact
justifies the appellation of ``Griffiths polynomials'' used in this paper for
$\cG_{i,j}(x,y)$.

In the following, we show that the Griffiths polynomials of $q$-Racah type have numerous
interesting properties.

Substituting the expression \eqref{eq:Tratpol} for $\cT_{j,a}(y,x;c_3,c_0,c_4,c_1;N)$
 into \eqref{eq:GpT},
one gets (for $x,y$ nonnegative integers and $x+y\leq N$):
\begin{align}\label{eq:Gpol}
    \cG_{i,j}(x,y)&=\Omega(i;c_1,c_2,c_3;N-j)\, \frac{(1-c_{0}c_4q^{2j+1})}{(q;q)_j}\,\frac{(qc_3;q)_y}{(qc_0;q)_y}
    \left( \frac{c_3c_4}{c_0c_1} \right)^{\frac12(N-y)}\,\left( q^{a-j}\frac{c_1}{c_4} \right)^{a}\,(q^{2j+1}c_0)^{\frac12N}
    \nonumber\\
   \times \sum_{a=0}^{N-j}& \frac{(q^{a-N};q)_j(qc_2;q)_a(qc_0;q)_{N-a}}{(q;q)_a(c_{0}c_{4}q^{j+1};q)_{N-a+1} (c_{1}c_{2}q^{a+1})_a(qc_1;q)_a}(q^{y-N};q)_a(q^{N-y+2}c_{1}c_{2}c_{4};q)_a   \nonumber\\
  &\times \pphi{4}{3}{q^{-i},\;c_{2}c_{3}q^{i+1}, \;q^{-a},\;c_{1}c_{2}q^{a+1}}{qc_2,\; q^{N-j+2}c_{1}c_{2}c_{3},\;q^{j-N}}{q;q}
          \pphi{4}{3}{q^{-a},\;c_{1}c_{2}q^{a+1}, \;q^{-x},\;c_{2}c_{4}q^{x+1}}{qc_2,\; q^{N-y+2}c_{1}c_{2}c_{4},\;q^{y-N}}{q;q}
  \nonumber\\
&\times\pphi{4}{3}{q^{j+a-N},\;q^{N-j+a+2}c_{1}c_{2}c_{3}, \;q^{y+a-N},\;q^{a+N-y+2}c_{1}c_{2}c_{4}}{c_{1}c_{2}q^{2a+2},\;q^{a-N}c_0^{-1},\;q^{a-N}}{q;q}\,.
\end{align}
Writing the basic hypergeometric functions appearing in \eqref{eq:Gpol} as finite sums, one can show that $\cG_{i,j}(x,y)$ divided by the first line of the r.h.s.~of \eqref{eq:Gpol} --- a function of the form $f(i,j)g(x,y)$ --- is a bivariate polynomial of total degree $N-j$ w.r.t. $\lambda(x;c_{2}c_{4})$ and $\lambda(y;c_{3}c_{0})$.

\subsection{Symmetry relations}\label{ssec:symm}
The Griffiths polynomials \eqref{eq:G3p} satisfy a symmetry relation \eqref{eq:symmgriff}
that links $\cG_{i,j}(x,y)$ to $\cG_{j,i}(y,x)$. This relation will play an important role
in the proof of the bispectrality of the Griffiths polynomials: put simply, the
bispectrality properties on $i$ and $x$ will be easily transposed to the bispectrality
properties on $j$ and $y$.

In order to prove this relation, we first need the following lemma.
\begin{lemm}\label{lemm:symm}
The Griffiths polynomials \eqref{eq:G3p} satisfy the following two relations:
\begin{align}\label{eq:Gxyyx}
\begin{aligned}
\cG_{i,j}(x,y;c_1,c_2,c_3,c_4;N) &=
\Big(\frac{c_0}{c_1}\Big)^{\!\frac i2}
\Big(\frac{c_3}{c_1}\Big)^{\!\frac j2}
\Big(\frac{c_1}{c_0}\Big)^{\!\frac x2}
\Big(\frac{c_1}{c_4}\Big)^{\!\frac y2}
\frac{\xi_i(c_2,c_3)\xi_j(c_0,c_4)}{\xi_x(c_2,c_4)\xi_y(c_0,c_3)}
\frac{\omega_{j}(c_0c_2c_4/c_1)}{\omega_i(c_3c_0c_2/c_1)} \,  \\
&\times\,\omega_{N}(qc_1c_2c_3)\cG_{i,j}(y,x;c_1^{-1},c_3^{-1},c_2^{-1},c_0^{-1};N)\Big|_{q \to q^{-1}} \,,
\end{aligned}
\end{align}
and
\begin{align}\label{eq:Gijji}
\begin{aligned}
\cG_{i,j}(x,y;c_1,c_2,c_3,c_4;N) &=
\Big(\frac{c_0}{c_1}\Big)^{\!\frac i2}
\Big(\frac{c_3}{c_1}\Big)^{\!\frac j2}
\Big(\frac{c_1}{c_0}\Big)^{\!\frac x2}
\Big(\frac{c_1}{c_4}\Big)^{\!\frac y2}
\frac{\xi_i(c_2,c_3)\xi_j(c_0,c_4)}{\xi_x(c_2,c_4)\xi_y(c_0,c_3)}
\frac{\omega_y(c_1c_3c_0/c_2)}{\omega_x(c_2c_4c_1/c_0)} \\
&\times \, \omega_{N}(qc_1c_2c_4) \,\cG_{j,i}(x,y;c_1^{-1},c_4^{-1},c_0^{-1},c_2^{-1};N)\Big|_{q \to q^{-1}} \,,
\end{aligned}
\end{align}
where $\xi$ and $\omega$ are given in \eqref{eq:omega}.
\end{lemm}
\proof
Rewrite the Griffiths polynomials \eqref{eq:G3p} as
\begin{align}\label{eq:symG1}
\begin{aligned}
\cG_{i,j}(x,y)=\sum_{b=0}^{N-j} \sum_{a=0}^{N-j}&\,
\frac{\omega_a({c_1c_2})}{\xi_a(c_1,c_2)}\,
\fp_i(a;c_1,c_2,c_3;N-j)\,
\delta_{a,b}\,
\\
& \times
\Big(\frac{c_3}{c_4}\Big)^{\!\frac b2}\,
\xi_b(c_1,c_2)\,
\fp_j(y;c_3,c_0,c_4;N-b)\,
\fp_b(x;c_4,c_2,c_1;N-y)\,
.
\end{aligned}
\end{align}
It is straightforward to check that \eqref{eq:symG1} reduces to \eqref{eq:G3p} upon
getting rid of the sum on $a$ with the Kronecker delta $\delta_{a,b}$ and using the
identities
\begin{align}\label{eq:usexo}
\xi_b(c_1,c_2)\,\xi_b(c_2,c_1)=1\,,\qquad
\omega_a(cd)=\omega_a(c)\,d^{\frac n2}\,.
\end{align}
Next, use the involution relation \eqref{eq:ortho2} to rewrite the Kronecker delta
$\delta_{a,b}$ in \eqref{eq:symG1}:
\begin{align}
\begin{aligned}
\cG_{i,j}(x,y)&=\sum_{b=0}^{N-j} \sum_{a=0}^{N-j} \sum_{d=0}^{N-j}
\frac{\omega_a({c_1c_2})}{\xi_a(c_1,c_2)}\,
\fp_i(a;c_1,c_2,c_3;N-j)\,
\fp_a(d;c_3,c_1,c_2;N-j)\,
\\ & \times
\Big(\frac{c_3}{c_4}\Big)^{\!\frac b2}\,
\xi_b(c_1,c_2)\,
\fp_d(b;c_2,c_1,c_3;N-j)\,
\fp_j(y;c_3,c_0,c_4;N-b)\,
\fp_b(x;c_4,c_2,c_1;N-y)\,
.
\end{aligned}
\end{align}
Since the bounds on $a$, $b$, $d$ are independent of each other, the sums can be
freely interchanged. Apply the Racah relation \eqref{eq:racahrel} on the first
two polynomials (sum over $a$). This leads to
\begin{align}\label{eq:tmp1}
\begin{aligned}
\cG_{i,j}(x,y)&=\sum_{b=0}^{N-j} \sum_{d=0}^{N-j}
\omega_{N-j}(qc_1c_2c_3)\,
\frac{\xi_i(c_2,c_3)\xi_d(c_3,c_1)}{\omega_i(c_2c_3)\omega_d(c_3c_1)}\,
\fp_i(d;c_1,c_3,c_2;N-j)\,
\\ & \times
\Big(\frac{c_3}{c_4}\Big)^{\!\frac b2}\,
\xi_b(c_1,c_2)\,
\fp_d(b;c_2,c_1,c_3;N-j)\,
\fp_j(y;c_3,c_0,c_4;N-b)\,
\fp_b(x;c_4,c_2,c_1;N-y)\,
.
\end{aligned}
\end{align}
Now use the Biedenharn--Elliot relation \eqref{eq:biedenrel} on the last three
polynomials (sum over $b$).
Hence one gets
\begin{align}\label{eq:tmp2}
\begin{aligned}
\cG_{i,j}(x,y)&=
\Big(\frac{c_3}{c_2}\Big)^{\!\frac j2}\,
\Big(\frac{c_0}{c_1}\Big)^{\!\frac x2}\,
\Big(\frac{c_1}{c_4}\Big)^{\!\frac y2}\,
\frac{\omega_{N-j}(qc_1c_2c_3)}{\omega_i(c_2c_3)}\,
\frac{\xi_i(c_2,c_3)\xi_j(c_0,c_4)}{\xi_x(c_2,c_4)\xi_y(c_0,c_3)}\,
\\
& \times \sum_{d=0}^{N-j}
\frac{1}{\omega_d(\frac{c_3c_1c_0}{c_2})}\,
\fp_i(d;c_1,c_3,c_2;N-j)\,
\fp_j(x;c_2,c_4,c_0;N-d)\,
\fp_d(y;c_0,c_3,c_1;N-x)\,
.
\end{aligned}
\end{align}
Using the symmetry relation \eqref{eq:symR} for the Racah polynomials and the equality
$\frac{1}{\omega_d(c)} = \omega_d(c^{-1})\big|_{q \to q^{-1}}$, the sum over $d$ in
\eqref{eq:tmp2} can now be written as Griffiths polynomials in the indeterminate
$q^{-1}$:
\begin{equation}
\begin{split}\label{eq:Ginv}
& \sum_{d=0}^{N-j}
\frac{1}{\omega_d(\frac{c_3c_1c_0}{c_2})}\,
\fp_i(d;c_1,c_3,c_2;N-j)\,
\fp_j(x;c_2,c_4,c_0;N-d)\,
\fp_d(y;c_0,c_3,c_1;N-x)\,
\\
& =
\Big(\frac{c_2}{c_1}\Big)^{\!j}\,
\Big(\frac{c_1}{c_0}\Big)^{\!x}\,
\sum_{d=0}^{N-j} \Big[
\omega_d\!\left(\!\frac{c_1^{-1}c_3^{-1}c_2^{-1}}{c_0^{-1}}\!\right)\,
\fp_i(d;c_1^{-1},c_3^{-1},c_2^{-1};N-j)\, \\
&\hspace{9em}\,\times
\fp_j(x;c_2^{-1},c_4^{-1},c_0^{-1};N-d)\,
\fp_d(y;c_0^{-1},c_3^{-1},c_1^{-1};N-x)\,
\Big]\Big|_{q \to q^{-1}} \\
& =
\Big(\frac{c_2}{c_1}\Big)^{\!j}\,
\Big(\frac{c_1}{c_0}\Big)^{\!x}\,
\cG_{i,j}(y,x;c_1^{-1},c_3^{-1},c_2^{-1},c_0^{-1};N)\Big|_{q \to q^{-1}}\,.
\end{split}
\end{equation}
Substituting this expression in \eqref{eq:tmp2} and expressing $\omega_{N-j}$ in term of
$\omega_N$ and $\omega_j$, one gets the result \eqref{eq:Gxyyx}.
The proof of relation \eqref{eq:Gijji} proceeds along the same lines.
\endproof
The previous lemma allows us to get a nice symmetry relation for the Griffiths polynomials
when the two variables and the two indices are exchanged simultaneously.
\begin{theo}\label{thm:symG}
The Griffiths polynomials \eqref{eq:G3p} satisfy the following symmetry relation:
\begin{align}
\label{eq:symmgriff}
\begin{aligned}
&
\cG_{i,j}(x,y;c_1,c_2,c_3,c_4;N)\
\omega_i\Big(\dfrac{c_2c_3c_4}{c_1}\Big)\,\omega_x\Big(\dfrac{c_4c_2c_1}{c_3}\Big)\,
c_0^{N/2} \;
\\
=\ &
\cG_{j,i}(y,x;c_1,c_0,c_4,c_3;N)\
\omega_j\Big(\dfrac{c_3c_0c_4}{c_1}\Big)\,\omega_y\Big(\dfrac{c_1c_0c_3}{c_4}\Big)\,
c_2^{N/2} \;
.
\end{aligned}
\end{align}
\end{theo}
\proof
Compare the two expressions of $\cG_{i,j}(x,y;c_1,c_2,c_3,c_4;N)$ given in Lemma
\ref{lemm:symm}. This yields a relation between
$\cG_{i,j}(y,x;c_1^{-1},c_3^{-1},c_2^{-1},c_0^{-1};N)\vert_{q \to q^{-1}}$ and
$\cG_{j,i}(x,y;c_1^{-1},c_4^{-1},c_0^{-1},c_2^{-1};N)\vert_{q \to q^{-1}}$. Replacing
$q^{-1} \mapsto q$ and $c_i^{-1} \mapsto c_i$ in this expression and relabeling the
$c_i$'s, one obtains the desired expression \eqref{eq:symmgriff}.
\endproof
\begin{rema}
In the limit $q\to1$ (with $c_i\mapsto q^{c_i}$), the symmetry relations given in Lemma
\ref{lemm:symm} and Theorem \ref{thm:symG} provide new symmetry relations for the
Griffiths polynomials of Racah type:
\begin{align}
G_{i,j}(x,y) &=
G_{i,j}(y,x;c_1,c_3,c_2,c_0;N)\
(-1)^{i+j+N}\,
\frac{(c_4+1)_x\,(c_3+1)_y\,(c_2+1)_i\,(c_0+1)_j}
     {(c_2+1)_x\,(c_0+1)_y\,(c_3+1)_i\,(c_4+1)_j}\,
,\label{eq:symRacah1}
\\
G_{i,j}(x,y) &=
G_{j,i}(x,y;c_1,c_4,c_0,c_2;N)\
(-1)^{x+y+N}\,
\frac{(c_4+1)_x\,(c_3+1)_y\,(c_2+1)_i\,(c_0+1)_j}
     {(c_2+1)_x\,(c_0+1)_y\,(c_3+1)_i\,(c_4+1)_j}\,
,
\\
G_{i,j}(x,y) &=
G_{j,i}(y,x;c_1,c_0,c_4,c_3;N)\
(-1)^{x+y+i+j}\,
,
\end{align}
where $(a)_n$ is the Pochhammer symbol, see Appendix \ref{App:def}.
These new relations may be used to simplify proofs given in \cite{GriffithR}.
\end{rema}

\subsection{Duality and biorthogonality}\label{ssec:db}
The Griffiths polynomials \eqref{eq:G3p} obey a duality relation that connects two
polynomials with a different set of parameters, as was the case for Tratnik polynomials.
\begin{prop}
The Griffiths polynomials $\cG_{i,j}$ obey the duality relation
\begin{equation}\label{eq:dual-griff}
\cG_{i,j}(x,y;c_1,c_2,c_3,c_4;N) =
\frac{\Omega(i;c_1,c_2,c_3;N-j)\,\Lambda(j;c_4,c_0,c_3,c_1;N)}
     {\Omega(x;c_1,c_2,c_4;N-y)\,\Lambda(y;c_3,c_0,c_4,c_1;N)}\,
\cG_{x,y}(i,j;c_1,c_2,c_4,c_3;N)
\end{equation}
where $\Omega$ and $\Lambda$ are given respectively by \eqref{def:Omega} and
\eqref{eq:Lambda}.
\end{prop}
\proof
This is a consequence of the duality relation \eqref{eq:dual} for the Racah polynomials
as well as the identities \eqref{eq:*} and \eqref{eq:**}.
\endproof
The Griffiths polynomials \eqref{eq:G3p} are not orthogonal; rather they are
\textit{biorthogonal}. Both functions that are orthogonal with one another are Griffiths
polynomials, but of different parameters.
\begin{prop}
The Griffiths polynomials are biorthogonal:
\begin{align}\label{eq:orthoG}
&\sum_{\atopn{x,y\geq0}{x+y\leq N}}
\frac{\Omega(x;c_1,c_2,c_4;N-y)\,\Lambda(y;c_3,c_0,c_4,c_1;N)}
     {\xi_x(c_2,c_4)\xi_y(c_0,c_3)(\frac{c_1}{c_0})^{x/2}\,(\frac{c_4}{c_1})^{y/2}}\,
\cG_{i,j}(x,y;c_1,c_2,c_3,c_4;N)
\,\cG_{k,\ell}(y,x;c_1,c_3,c_2,c_0;N)
\nonumber\\
&\qquad =
\delta_{i,k}\,\delta_{j,\ell}\,
\frac{\omega_j(\frac{c_2c_0c_4}{c_3})\,\omega_{N}(qc_1c_2c_3)}{\omega_i({c_2c_3})}\,
\frac{\Omega(i;c_1,c_2,c_3;N-j)\,\Lambda(j;c_4,c_0,c_3,c_1;N)}
     {\xi_i(c_2,c_3)\,{\xi_j(c_0,c_4)}}\,.
\end{align}
The functions $\xi$ and $\omega$ are given in \eqref{eq:omega}.
\end{prop}
\proof
Consider the sum
\begin{equation}\label{eq:start}
\sum_{\atopn{x,y\geq0}{x+y\leq N}}
\frac{\Omega(x;c_1,c_2,c_4;N-y)\,\Lambda(y;c_3,c_0,c_4,c_1;N)}
     {\Omega(i;c_1,c_2,c_3;N-j)\,\Lambda(j;c_4,c_0,c_3,c_1;N)}\,
\cG_{i,j}(x,y;c_1,c_2,c_3,c_4;N)
\,\tilde\cG_{k,\ell}(x,y)\,,
\end{equation}
where
\begin{equation}
\tilde\cG_{k,\ell}(x,y) =\sum_{b=0}^{N-\ell}
\frac{\fp_k(b;c_1,c_2,c_3;N-\ell)\,\fp_\ell(y;c_3,c_0,c_4;N-b)\,\fp_b(x;c_4,c_2,c_1;N-y)}
     {\omega_b(\frac{c_1c_2c_4}{c_3})}\,.
\end{equation}
Expressing $\cG_{i,j}$ and $\tilde\cG_{k,\ell}$ in term of univariate $q$-Racah
polynomials $\fp_n$, \eqref{eq:start} is rewritten as a quadruple sum
(on $x$, $y$, $a$ and $b$)
of univariate polynomials with the following coefficient entering the sum:
\begin{equation}\label{coef}
\frac{\Omega(x;c_1,c_2,c_4;N-y)\,\Lambda(y;c_3,c_0,c_4,c_1;N)}
     {\Omega(i;c_1,c_2,c_3;N-y)\,\Lambda(j;c_4,c_0,c_3,c_1;N)}
\frac{\omega_a(\frac{c_1c_2c_3}{c_4})}{\omega_b(\frac{c_1c_2c_4}{c_3})}\,.
\end{equation}
Using relation \eqref{eq:*} to rewrite each of the $\Lambda$ functions,
the coefficient \eqref{coef} is reexpressed as:
\begin{equation}\label{coef2}
\frac{\Omega(x;c_1,c_2,c_4;N-y)\,\Omega(a;c_3,c_2,c_1;N-j)\,\Omega(y;c_4,c_0,c_3;N-a)}
     {\Omega(a;c_4,c_2,c_1;N-y)\,\Omega(i;c_1,c_2,c_3;N-j)\,\Omega(j;c_3,c_0,c_4;N-a)}
\frac{\Lambda(a;c_1,c_2,c_4,c_3;N)\omega_a(\frac{c_1c_2c_3}{c_4})}
     {\Lambda(a;c_1,c_2,c_3,c_4;N)\omega_b(\frac{c_1c_2c_4}{c_3})}\,.
\end{equation}
Now, starting from \eqref{eq:start}, perform the sum on $x$ and use the orthogonality
relation \eqref{eq:ortho} for $q$-Racah polynomials to get a $\delta_{a,b}$ term that
lets one evaluate the sum on $b$.
By virtue of relation \eqref{eq:*}, the term
$\frac{\Lambda(a;c_1,c_2,c_4,c_3;N)\,\omega_a({c_1c_2c_3}/{c_4})}
      {\Lambda(a;c_1,c_2,c_3,c_4;N)\,\omega_b({c_1c_2c_4}/{c_3})}$ reduces to $1$.
Then, perform the sum on $y$ to obtain a $\delta_{j,\ell}$ term through the
orthogonality relation. Lastly, evaluate the sum on $a$, again using the orthogonality
relation, to obtain another $\delta_{i,k}$ term. Hence, the expression \eqref{eq:start} is
$\delta_{j,\ell}\,\delta_{i,k}$.
Next, notice that equation \eqref{eq:Ginv} shows that
\begin{equation}
\tilde\cG_{k,\ell}(x,y) =
\Big(\frac{c_3}{c_1}\Big)^{\!\ell}\,
\Big(\frac{c_1}{c_4}\Big)^{\!y}\,
\cG_{k,\ell}(x,y;c_1^{-1},c_2^{-1},c_3^{-1},c_4^{-1};N)\Big|_{q \to q^{-1}}\,.
\end{equation}
Using the relation \eqref{eq:Gxyyx} and the property
$\xi_n(c,c')=1/\xi_n(c',c)\,,$
the final result \eqref{eq:orthoG} is obtained.
\endproof
Let us stress that in the limit $q\to 1$, the biorthogonality relation becomes an
orthogonality relation, in accordance with the results obtained in \cite{GriffithR}.
The symmetry relation \eqref{eq:symRacah1} is used to obtain this result.

Note also that the Griffiths polynomial
$\cG_{i,j}(x,y;c_1,c_2,c_2,\pm\frac{q^{-N-1}}{c_2\sqrt{qc_1}};N)$ is orthogonal.
\begin{coro}
The Griffiths polynomials $\cG_{i,j}$ obey an involution relation
\begin{align}\label{eq:invol}
\begin{aligned}
&\sum_{\atopn{x,y\geq0}{x+y\leq N}}
\Big(\frac{c_1}{c_0}\Big)^{\!\frac x2}\,
\Big(\frac{c_4}{c_1}\Big)^{\!\frac y2}\,
\xi_x(c_2,c_4)\,
\xi_y(c_0,c_3)\,
\cG_{i,j}(x,y;c_1,c_2,c_3,c_4;N)\,
\cG_{y,x}(k,\ell;c_1,c_3,c_0,c_2;N)\,
\\
&\qquad =
\delta_{i,k}\,\delta_{j,\ell}\
\omega_{N}(qc_1c_2c_3)\,
\frac{\omega_j(\frac{c_3c_0c_4}{c_2})}{\omega_i({c_2c_3})} \,
\xi_i(c_2,c_3)\,
{\xi_j(c_0,c_4)} \,.
\end{aligned}
\end{align}
\end{coro}
\proof
As a direct consequence of the duality and biorthogonality relations, one gets:
\begin{align}
&\sum_{\atopn{x,y\geq0}{x+y\leq N}}
\frac{\Omega(x;c_1,c_2,c_4;N-y)\,\Lambda(y;c_3,c_0,c_4,c_1;N)}
     {\Omega(y;c_1,c_3,c_0;N-x)\,\Lambda(x;c_2,c_4,c_0,c_1;N)}\,
\frac{\cG_{i,j}(x,y;c_1,c_2,c_3,c_4;N)\,\cG_{y,x}(k,\ell;c_1,c_3,c_0,c_2;N)}
     {\xi_x(c_2,c_4)(\frac{c_1}{c_0})^{x/2}\,\xi_y(c_0,c_3)(\frac{c_4}{c_1})^{y/2}}
\nonumber\\
&\qquad =
\frac{\omega_j(\frac{c_2c_0c_4}{c_3})\,\omega_{N}(qc_1c_2c_3)}{\omega_i({c_2c_3})}\,
\frac{\Omega(i;c_1,c_2,c_3;N-j)\,\Lambda(j;c_4,c_0,c_3,c_1;N)}
     {\Omega(i;c_1,c_3,c_2;N-j)\,\Lambda(j;c_0,c_4,c_2,c_1;N)}\,
\frac{\delta_{i,k}\,\delta_{j,\ell}}{\xi_i(c_2,c_3)\,{\xi_j(c_0,c_4)}}\,.
\end{align}
Then using the explicit expressions of the functions $\Omega$ and $\Lambda$,
the result \eqref{eq:invol} follows.
\endproof

\subsection{Bispectral property}\label{ssec:bispectralG}
In this subsection, we obtain the bispectral property of the Griffiths polynomials of
$q$-Racah type.
\begin{prop}\label{prop:rec-griff}
The Griffiths polynomials of $q$-Racah type satisfy the following two recurrence
relations:
\begin{align}
\label{eq:rec1-griff}
\lambda(y;c_3c_0)\,\cG_{i,j}(x,y) &= \sum_{\eps,\eps'=0,\pm1}
\,A^{\eps,\eps'}_{i+\eps,j+\eps'}(c_1,c_2,c_3,c_4;N)\,\cG_{i+\eps,j+\eps'}(x,y)\,,\\
\label{eq:rec2-griff}
\lambda(x;c_4c_2) \,\cG_{i,j}(x,y) &=\sum_{\eps,\eps'=0,\pm1}
\frac{\omega_{\eps}(q^{2i}c_2c_3c_4/c_1)}{\omega_{\eps'}(q^{2j}c_0c_3c_4/c_1)} \,
A^{\eps',\eps}_{j+\eps',i+\eps}(c_1,c_0,c_4,c_3;N)\,\cG_{i+\eps,j+\eps'}(x,y)\,,
\end{align}
where $A^{\eps,\eps'}_{i,j}$ are given in \eqref{def:Aij} and $\omega_\eps$ in
\eqref{eq:omega}.
\end{prop}
\proof
The proof of \eqref{eq:rec1-griff} consists in writing the Griffiths polynomials using the
form \eqref{eq:GTp} and in using the recurrence relation \eqref{eq:rec2-trat} of the
Tratnik polynomials.

To simplify the calculation, we define
\begin{equation}
\rho_{i,j} =
\frac{\omega_j\Big(\dfrac{c_3c_0c_4}{c_1}\Big)\,\omega_y\Big(\dfrac{c_1c_0c_3}{c_4}\Big)}
     {\omega_i\Big(\dfrac{c_2c_3c_4}{c_1}\Big)\,\omega_x\Big(\dfrac{c_4c_2c_1}{c_3}\Big)}
\left(\dfrac{c_2}{c_0}\right)^{N/2}\,.
\end{equation}
To prove \eqref{eq:rec2-griff}, rewrite its l.h.s.~using the symmetry relation
\eqref{eq:symmgriff} and apply the first recurrence relation \eqref{eq:rec1-griff}
(up to a suitable change of variables and indices):
\begin{align}\label{eq:proofrec2griff}
\begin{aligned}
&\lambda(x;c_2c_{4}) \,\cG_{i,j}(x,y;c_1,c_2,c_3,c_4;N) =
\rho_{i,j}\, \lambda(x;c_4c_2) \,\cG_{j,i}(y,x;c_1,c_0,c_4,c_3;N) \\
&\quad= \rho_{i,j} \sum_{\eps,\eps'=0,\pm1}\,
A^{\eps',\eps}_{j+\eps',i+\eps}(c_1,c_0,c_4,c_3;N)\,
\cG_{j+\eps',i+\eps}(y,x;c_1,c_0,c_4,c_3;N)\\
&\quad= \sum_{\eps,\eps'=0,\pm1} \frac{\rho_{i,j}}{\rho_{i+\eps,j+\eps'}}\,
A^{\eps',\eps}_{j+\eps',i+\eps}(c_1,c_0,c_4,c_3;N)\,
\cG_{i+\eps,j+\eps'}(x,y;c_1,c_2,c_3,c_4;N)\,.
\end{aligned}
\end{align}
Examine the last line of \eqref{eq:proofrec2griff}; the r.h.s.~of relation
\eqref{eq:rec2-griff} is recognized by remarking that the ratio of the functions $\rho$ is
given by
\begin{equation}
\frac{\rho_{i,j}}{\rho_{i+\eps,j+\eps'}} =
\frac{\omega_{i+\eps}\Big(\dfrac{c_2c_3c_4}{c_1}\Big)\,
      \omega_j\Big(\dfrac{c_3c_0c_4}{c_1}\Big)}
     {\omega_i\Big(\dfrac{c_2c_3c_4}{c_1}\Big)\,
      \omega_{j+\eps'}\Big(\dfrac{c_3c_0c_4}{c_1}\Big)}
= \frac{\omega_{\eps}(q^{2i}c_2c_3c_4/c_1)}{\omega_{\eps'}(q^{2j}c_0c_3c_4/c_1)}\,,
\end{equation}
which concludes the proof.
\endproof
\begin{prop}
The Griffiths polynomials of $q$-Racah type satisfy the following two difference
relations:
\begin{align}
\label{eq:diff1-griff}
\lambda(j;c_0c_4)\,\cG_{i,j}(x,y) &= \sum_{\eps,\eps'=0,\pm1} \,
A^{\eps,\eps'}_{x,y}(c_1,c_2,c_4,c_3;N)\,\cG_{i,j}(x-\eps,y-\eps')\,,\\
\label{eq:diff2-griff}
\lambda(i;c_2c_3) \,\cG_{i,j}(x,y) &=\sum_{\eps,\eps'=0,\pm1}
\frac{\omega_{-\eps'}(q^{2x}c_1c_2c_4/c_3)}{\omega_{-\eps}(q^{2y}c_1c_0c_3/c_4)} \,
A^{\eps',\eps}_{y,x}(c_1,c_0,c_3,c_4;N)\,\cG_{i,j}(x-\eps',y-\eps)\,,
\end{align}
where $A^{\eps,\eps'}_{x,y}$ are given in \eqref{def:Aij} and $\omega_\eps$ in
\eqref{eq:omega}.
\end{prop}
\proof
The proof can be done mimicking the one of Proposition \ref{prop:rec-griff}.
Alternatively, one can use the duality property \eqref{eq:dual-griff} of the Griffiths
polynomials to obtain it from the recurrence relations.
\endproof
It has been shown in \cite{Zhe2} that $\aw$ is the bispectral algebra of the $q$-Racah
polynomials. For the bivariate Griffiths polynomials of $q$-Racah type introduced here,
the bispectral algebra is the rank $2$ Askey--Wilson algebra $\haw$ defined in Appendix
\ref{app:def-aw4}.

\begin{rema}[\textbf{Generalized Favard theorem}]
We show that the conditions of the generalized Favard theorem (see
Remark \ref{rem:Favard-T}) are not satisfied for the Griffiths polynomials, which is
consistent with the fact that they are biorthogonal.
For nonnegative integers $i,j$ such that $i+j\leq N$, one defines
\begin{align}\label{eq:Tbar}
\overline\cG_{i,j}(x,y)&=\cG_{i,j}(x,y)\,/\,\cG_{N,0}(x,y)
\end{align}
where $\cG_{i,j}(x,y)$ is given by \eqref{eq:Gpol}.
It is of total degree $N-i$ w.r.t. $\lambda(x;c_4c_2)$ and $\lambda(y;c_0c_3)$.
Consider
\begin{equation}\label{eq:rec-FavardG}
\GG_{n}=\Big(\chi_{n,0}\overline\cG_{N-n,0}\,,\ \chi_{n,1}\overline\cG_{N-n,1}\,,\
\dots\,,\ \chi_{n,n}\overline\cG_{N-n,n}\Big)^T
\end{equation}
and look for coefficients $\chi$ such that the recurrence relations
\eqref{eq:rec1-griff} and \eqref{eq:rec2-griff} can be put in the form of
\eqref{eq:rec-FavardG}. This leads to the constraints
\begin{align}\label{eq:cont-chi1}
\begin{aligned}
\left(\frac{\chi_{i,j+1}}{\chi_{i,j}}\right)^2 &=
\frac{A^{0,+}_{N-i,j+1}(c_1,c_2,c_3,c_4;N)}{A^{0,-}_{N-i,j}(c_1,c_2,c_3,c_4;N)}\,,&\quad
\left(\frac{\chi_{i+1,j}}{\chi_{i,j}}\right)^2 &=
\frac{A^{-,0}_{N-i-1,j}(c_1,c_2,c_3,c_4;N)}{A^{+,0}_{N-i,j}(c_1,c_2,c_3,c_4;N)}\,,\\
\left(\frac{\chi_{i+1,j+1}}{\chi_{i,j}}\right)^2 &=
\frac{A^{-,+}_{N-i-1,j+1}(c_1,c_2,c_3,c_4;N)}{A^{+,-}_{N-i,j}(c_1,c_2,c_3,c_4;N)}\,,&\quad
\left(\frac{\chi_{i,j+1}}{\chi_{i+1,j}}\right)^2 &=
\frac{A^{+,+}_{N-i,j+1}(c_1,c_2,c_3,c_4;N)}{A^{-,-}_{N-i-1,j}(c_1,c_2,c_3,c_4;N)}\,,
\end{aligned}
\end{align}
and
\begin{align}\label{eq:cont-chi2}
\begin{aligned}
\left(\frac{\chi_{i,j+1}}{\chi_{i,j}}\right)^2 &=
\frac{\omega_{-}(q^{2j+2}c_3c_0c_4/c_1)}{\omega_{+}(q^{2j}c_3c_0c_4/c_1)}\,
\,\frac{A^{+,0}_{j+1,N-i}(c_1,c_0,c_4,c_3;N)}{A^{-,0}_{j,N-i}(c_1,c_0,c_4,c_3;N)}\,,\\
\left(\frac{\chi_{i+1,j}}{\chi_{i,j}}\right)^2 &=
\frac{\omega_{-}(q^{2(N-i)}c_2c_3c_4/c_1)}{\omega_{+}(q^{2(N-i-1)}c_2c_3c_4/c_1)}\,
\,\frac{A^{0,-}_{j,N-i-1}(c_1,c_0,c_4,c_3;N)}{A^{0,+}_{j,N-i}(c_1,c_0,c_4,c_3;N)}\,,\\
\left(\frac{\chi_{i+1,j+1}}{\chi_{i,j}}\right)^2 &=
\frac{\omega_{-}(q^{2(N-i)}c_2c_3c_4/c_1)}{\omega_{+}(q^{2j+2}c_3c_0c_4/c_1)}\,
\frac{\omega_{-}(q^{2j}c_3c_0c_4/c_1)}{\omega_{+}(q^{2(N-i-1)}c_2c_3c_4/c_1)}\,
\frac{A^{+,-}_{j+1,N-i-1}(c_1,c_0,c_4,c_3;N)}{A^{-,+}_{j,N-i}(c_1,c_0,c_4,c_3;N)}\,,\\
\left(\frac{\chi_{i,j+1}}{\chi_{i+1,j}}\right)^2 &=
\frac{\omega_{+}(q^{2(N-i-1)}c_2c_3c_4/c_1)}{\omega_{+}(q^{2j}c_3c_0c_4/c_1)}\,
\frac{\omega_{-}(q^{2j+2}c_3c_0c_4/c_1)}{\omega_{-}(q^{2(N-i)}c_2c_3c_4/c_1)}\,
\frac{A^{+,+}_{j+1,N-i}(c_1,c_0,c_4,c_3;N)}{A^{-,-}_{j,N-i-1}(c_1,c_0,c_4,c_3;N)}\,.
\end{aligned}
\end{align}
From the calculations done for Tratnik polynomials, one knows that in
\eqref{eq:cont-chi1}, the second line is a consequence of the first line. Similarly, the
last two lines of \eqref{eq:cont-chi2} are a consequence of the first two ones.
However, the equations \eqref{eq:cont-chi1} and \eqref{eq:cont-chi2} are inconsistent.
For instance, the ratio of both expressions of
$\big(\tfrac{\chi_{i,j+1}}{\chi_{i,j}}\big)^2$ given in \eqref{eq:cont-chi1} and
\eqref{eq:cont-chi2} is equal to
\begin{equation}
\frac{q^{2j-2N-1}}{c_1c_2c_3}\neq1\,.
\end{equation}
Thus, the Griffiths polynomials of $q$-Racah type cannot be orthogonal. Note that in the limit
$c_i\mapsto q^{c_i}$, $q\to1$, the expressions become consistent,
in accordance with the orthogonality property of the Griffiths polynomials of Racah type
\cite{GriffithR}.
\end{rema}

\section{Outlook}\label{sec:conclu}
This paper is devoted to the introduction and the study of new bivariate polynomials, the
\textit{Griffiths polynomials of $q$-Racah type}.
From these polynomials, one could perform various limits, so as to get
bivariate functions involving other monovariate polynomials such as the $q$-Hahn or the
$q$-Krawtchouk polynomials.
These limits could have interesting uses;
for example, the bivariate Tratnik polynomials of $q$-Krawtchouk type appear naturally in
the study of $q$-rotations \cite{GPV1} and in the representation theory of $SU_q(3)$
\cite{BKV} and hence the obtained polynomials may generalize these results.

The orthogonality property of these polynomials also deserves some studies. Unlike the
Griffiths polynomials of Racah type, which are orthogonal, the polynomials introduced in
this paper are biorthogonal. Clearly, some restriction on the parameter space can lead to
orthogonal polynomials. However, it could be that other constructions lead to truly
orthogonal polynomials; these polynomials would potentially possess a higher number of
parameters. In that case, it
is likely that the underlying algebraic structure would be different from the
Askey--Wilson algebra $\mathbf{aw}(4)$. The existence of this new algebra remains to be
elucidated.

A natural generalization of the results obtained here would be to consider multivariate
polynomials in more than two variables.
These $n$ variable polynomials would be a generalization of the $n$ variable Tratnik
polynomials introduced and studied in \cite{GR2005,Iliev}.
Their bispectral property should provide a representation of the higher rank Askey--Wilson
algebra $\mathbf{aw}(n+2)$ which has been defined in \cite{awn}. It would also be
interesting to study the representation theory of this algebra on its own to obtain
results on these polynomials as a byproduct.

The Leonard theorem has initially been proved in order to try to classify the so-called
$P$- and $Q$-polynomial schemes or distance regular graph (see \cite{Leo,BI}).
These different notions have been generalized recently in \cite{BCPVZ,BKZZ,BCVZZ}.
It would be interesting to generalize the Leonard theorem to the multivariate case and to
obtain a classification of bispectral multivariate polynomials.

\paragraph{Acknowledgements:}
The authors warmly thank Luc Vinet and Meri Zaimi for stimulating discussions.
N.~Cramp\'e is partially supported by the international research project AAPT of the CNRS.
L.~Frappat and E.~Ragoucy are partially supported by Universit\'e Savoie Mont Blanc and
Conseil Savoie Mont Blanc grant POSITIPh.
J.~Gaboriaud is supported by JSPS KAKENHI Grant Numbers 22F21320 and 22KF0189.

\appendix
\section{Basic hypergeometric functions}\label{App:def}
The basic hypergeometric functions, or $q$-hypergeometric functions, are defined by the
following series, for $r$ and $s$ nonnegative integers:
\begin{equation}\label{eq:defphi}
\pphi{r}{s}{a_1,\;a_2,\;\dots,\;a_r}{b_1,\;b_2,\dots,\;b_s}{q;z}=
\sum_{k=0}^{+\infty} \frac{(a_1,a_2,\dots,a_r;q)_k}{(b_1,b_2,\dots,b_s;q)_k} (
-1)^{(1+s-r)k} q^{(1+s-r)\binom{k}{2}}\frac{z^k}{(q;q)_k}\,,
\end{equation}
where
\begin{align}
(a_1,a_2,\dots,a_r;q)_k=(a_1;q)_k(a_2;q)_k\dots (a_r;q)_k\,,
\end{align}
and the $q$-Pochhammer symbol $(a;q)_n$ is defined as
\begin{align}
(a;q)_0=1\,,\qquad
(a;q)_k=\prod_{\ell=0}^{k-1}(1-aq^\ell)\,,\quad k=1,2,3,\dots.
\end{align}
The $q$-Pochhammer symbol satisfies the following transformation formula:
\begin{equation}\label{eq:symPoch}
(\tfrac{1}{a};q)_{n} = (-\nicefrac{1}{a})^n\,
q^{\binom{n}{2}}\,({a}\,q^{-n+1};q)_n\,.
\end{equation}
When $a_1=q^{-n}$ for some nonnegative integer $n$, the series in \eqref{eq:defphi}
truncates and becomes a sum over $k$ from $0$ to $n$.
Let us emphasize that the $q$-hypergeometric functions are invariant under any permutation
either of its parameters $a_1,a_2,\dots,a_r$ or of its parameters $b_1,b_2,\dots,b_s$.
It also satisfies
\begin{align}\label{eq:phiqinv}
\pphi{4}{3}{q^{n},\;a_2, \;a_3,\; a_4 }{b_1,\;b_2,\;b_3}{q^{-1};q^{-1}}=
\pphi{4}{3}{q^{-n},\; a_2^{-1},\; a_3^{-1},\; a_4^{-1}}
           {b_1^{-1},\; b_2^{-1},\; b_3^{-1}}{q;\frac{q^na_2a_3a_4}{b_1b_2b_3}}\,.
\end{align}

\subsubsection*{Limit $q\to1$.}
In the limit $q\to1$, the $q$-generalized hypergeometric function
$\pphi{4}{3}{q^{-n},\; q^{a_2}, \;q^{a_3},\; q^{a_4}}{q^{b_1},\; q^{b_2},\; q^{b_3}}{q;q}$
tends to the generalized hypergeometric function
$\pFq{4}{3}{{-n},\; {a_2},\; {a_3},\; {a_4}}{{b_1},\; {b_2},\; {b_3}}{1}$ and
the renormalized $q$-Pochhammer symbol tends to the Pochhammer symbol:
\begin{equation}
\lim_{q\to1}\frac{(q^a;q)_k}{(q;q)_k} = \prod_{n=0}^{k-1}(a+n) = (a)_k \,.
\end{equation}

\section{Representations of the Askey--Wilson algebra $\aw$}\label{app:aw3}
In \cite{Zhe2}, it has been shown that the operators defined by the bispectral property of
the $q$-Racah polynomials satisfy an algebra, called the \textit{Askey--Wilson algebra}.
The Askey--Wilson algebra has subsequently found applications in various domains:
association schemes \cite{Leo,BI,Ter2}, Leonard pairs \cite{TV,Ter},
Kauffman bracket skein algebras \cite{CL,Skein5},
the centralizer of $U_q(sl_2)$ in tensor products of any three possibly different spin
representations \cite{CVZ2,CPVZ}
and symmetry of physical models \cite{KKM,Post,GVZ}.
In addition, its representation theory has been studied in \cite{Hua2}.
For more details about the Askey--Wilson algebra $\aw$, we refer to the review
\cite{avatar}.

\subsection{Defining relations for $\aw$}\label{}
We first provide the defining relations satisfied by the $\aw$ algebra.
We then exhibit a particular symmetry under which these relations are unchanged
and which we will use to obtain multiple representations in the next subsection.

\begin{defi}
The Askey--Wilson algebra $\aw$ is an unital algebra generated by
$C_{12}$, $C_{23}$ and $C_{13}$, with four central charges
$C_1$, $C_2$, $C_3$ and $C_{123}$ and a parameter $\scq$.
The generators obey the following three relations:
\begin{subequations}\label{eq:aw3def}
\begin{align}
\label{eq:aw3_i}
\frac{1}{\scq^2-\scq^{-2}}[C_{12},C_{23}]_\scq =
- C_{13}+\frac{1}{\scq+\scq^{-1}}(C_{3}C_{1}+C_{2}C_{123})\,,\\
\label{eq:aw3_ii}
\frac{1}{\scq^2-\scq^{-2}}[C_{23},C_{13}]_\scq =
- C_{12}+\frac{1}{\scq+\scq^{-1}}(C_{1}C_{2}+C_{3}C_{123})\,,\\
\label{eq:aw3_iii}
\frac{1}{\scq^2-\scq^{-2}}[C_{13},C_{12}]_\scq =
- C_{23}+\frac{1}{\scq+\scq^{-1}}(C_{2}C_{3}+C_{1}C_{123})\,,
\end{align}
\end{subequations}
where $[X,Y]_\scq=\scq XY-\scq^{-1}YX$.
\end{defi}
\begin{rema}\label{rema:aw3Z3}
The Askey--Wilson algebra $\aw$ has a $\ZZ_3$ symmetry.
In particular, the relations of the Askey--Wilson algebra $\aw$
are unchanged under the following cyclical permutations of the generators:
\begin{align}\label{}
\sigma~:\quad
\begin{aligned}
C_1&\mapsto C_2\,,\\
C_2&\mapsto C_3\,,\\
C_3&\mapsto C_1\,,
\end{aligned}
\qquad
C_{123}\mapsto C_{123}\,,
\qquad
\begin{aligned}
C_{12}&\mapsto C_{23}\,,\\
C_{23}&\mapsto C_{13}\,,\\
C_{13}&\mapsto C_{12}\,.
\end{aligned}
\end{align}
It is immediate to see that $\sigma^3=1$.
\end{rema}

\subsection{Representations of the algebra $\aw$}\label{sec:aw3reps}
We now provide $(N+1)$-dimensional representations of the Askey--Wilson algebra.

Define the functions $\xi$ and $\omega$ as
\begin{align}\label{eq:omega}
\xi_n(c_1,c_2)=\frac{ (qc_1;q)_n}{(qc_2;q)_n} \left( \frac{c_2}{c_1} \right)^{n/2}
\qquad \text{and} \qquad
\omega_n(c)&= (-1)^n  c^{n/2}\, q^{\binom{n+1}{2}}\,,
\end{align}
and let $\un$ denote the identity matrix of dimension $N+1$.

\paragraph{The $\cA$ matrices.}
Introduce the $(N+1)\times(N+1)$-dimensional matrices
$\cA_{12}$, $\cA_{23}$ and $\cA_{13}$.
Their non-vanishing entries are given as follows, for $0\leq i \leq N$:
\begin{align}\label{eq:awA}
\begin{aligned}
(\cA_{12})_{i,i}&=\lambda(i;c_1c_2)\,,\\
(\cA_{23})_{i-\eps,i}&=\Phi^{\eps}(i;c_3,c_2,c_1;N)\,,\\
(\cA_{13})_{i-\eps,i}&=
\frac{\xi_{i-\eps}(c_2,c_1)\omega_{i-\eps}(c_2c_1)}
{\xi_{i}(c_2,c_1)\omega_{i}(c_2c_1)}
\Phi^{\eps}(i;c_3,c_1,c_2;N)\,,
\end{aligned}
\hspace{4em}\eps\in\{0,\pm\}\,,
\end{align}
In the above and hereinafter, by convention the entries $(-1,i)$ and $(N+1,i)$ of
any of the $(N+1)\times(N+1)$-dimensional matrices are null,
\textit{e.g.}~$(\cA_{12})_{-1,0}=(\cA_{13})_{N+1,N}=0$.
The matrix $\cA_{12}$ is diagonal and both the matrices $\cA_{23}$ and $\cA_{13}$ are
tridiagonal matrices.

One observes that the matrices $\cA_{12}$ and $\cA_{23}$ encode respectively the
recurrence \eqref{eq:rec-rac} and difference \eqref{eq:rec-rac4} relations
of the $q$-Racah polynomials, in other words their bispectral property.
It is a standard result that the algebraic relations between the matrices that encode the
bispectral property of the $q$-Racah polynomials are those of the Askey--Wilson algebra
\cite{Zhe2}.
The explicit correspondence with the relations \eqref{eq:aw3def}
will be given in the later Proposition \eqref{prop:aw3rep}.

\paragraph{The $\cB$ matrices.}
The $(N+1)\times(N+1)$-dimensional matrices $\cB_{12}$, $\cB_{23}$ and $\cB_{13}$,
have the following non-vanishing entries, for $0\leq i \leq N$:
\begin{align}\label{eq:awB}
\begin{aligned}
(\cB_{23})_{i,i}&=\lambda(i;c_2c_3)\,,\\
(\cB_{13})_{i-\eps,i}&=\Phi^{\eps}(i;c_1,c_3,c_2;N)\,,\\
(\cB_{12})_{i-\eps,i}&=
\frac{\xi_{i-\eps}(c_3,c_2)\omega_{i-\eps}(c_3c_2)}
{\xi_{i}(c_3,c_2)\omega_{i}(c_3c_2)}
\Phi^{\eps}(i;c_1,c_2,c_3;N)\,,
\end{aligned}
\hspace{4em}\eps\in\{0,\pm\}\,,
\end{align}
with the functions $\xi$ and $\omega$ defined in \eqref{eq:omega}.

\paragraph{The $\cC$ matrices.}
The $(N+1)\times(N+1)$-dimensional matrices $\cC_{12}$, $\cC_{23}$ and $\cC_{13}$,
have the following non-vanishing entries, for $0\leq i \leq N$:
\begin{align}\label{eq:awC}
\begin{aligned}
(\cC_{13})_{i,i}&=\lambda(i;c_1c_3)\,,\\
(\cC_{12})_{i-\eps,i}&=\Phi^{\eps}(i;c_2,c_1,c_3;N)\,,\\
(\cC_{23})_{i-\eps,i}&=
\frac{\xi_{i-\eps}(c_1,c_3)\omega_{i-\eps}(c_1c_3)}
{\xi_{i}(c_1,c_3)\omega_{i}(c_1c_3)}
\Phi^{\eps}(i;c_2,c_3,c_1;N)\,,
\end{aligned}
\hspace{4em}\eps\in\{0,\pm\}\,,
\end{align}
with the functions $\xi$ and $\omega$ defined in \eqref{eq:omega}.

\begin{prop}
\label{prop:aw3rep}
Each set of $\cA$, $\cB$ and $\cC$ matrices
provides $\mbox{(N+1)}$-dimensional representations
of the Askey--Wilson algebra $\aw$.
Concretely, when the parameter $\scq$ of $\aw$ is equal to $\sqrt{q}$,
the following map is an algebra homomorphism:
\begin{equation}\label{eq:repaw}
\begin{array}{lcll}
C_i & \mapsto & (\sqrt{c_i}+\frac{1}{\sqrt{c_i}})\un \,,
&
\text{for }\ i=1,2,3\,,
\\[1.ex]
C_{ij} & \mapsto & \frac{1}{\sqrt{qc_ic_j}}\cX_{ij}
                  +(\sqrt{qc_ic_j}+\frac{1}{\sqrt{qc_ic_j}})\un\,,
\qquad
&
\text{for }\ ij=12,23,13\,,
\\[1.ex]
C_{123} & \mapsto & ( \sqrt{q^{2N+2}c_1c_2c_3}+\frac{1}{\sqrt{q^{2N+2}c_1c_2c_3}})\un\,.
\end{array}
\end{equation}
In the above, $\cX$ is chosen as either $\cA$, $\cB$, or $\cC$.
We will refer to these three representations as the
{$\cA$, $\cB$, and $\cC$ representations}.
\end{prop}
\proof
The algebraic relations can be checked by explicit matrix computations
or using symbolic manipulation software such as Mathematica or Maple.
\endproof
Due to the constraints \eqref{eq:cons1}, \eqref{eq:cons3} and \eqref{eq:cons2}, the
above three representations $\cA$, $\cB$, $\cC$ are irreducible.
Indeed, remarking that the spectrum of $\cA_{12}$ is simple and that $\cA_{23}$ is a
tridiagonal matrix where no off-diagonal entry vanishes allow to prove that the
representation is irreducible. The result on $\cB$ and $\cC$ then follows immediately
by symmetry.

As is well-known, the previous properties show that $(\cA_{12},\cA_{23})$ is
a realization of a Leonard pair \cite{Ter}.
Let us mention that the constraints
\eqref{eq:cons1}, \eqref{eq:cons3} and \eqref{eq:cons2} are necessary in order
to get a Leonard pair, due to the irreducibility condition entering
the definition of Leonard pairs.

\begin{rema}\label{rem:Z3}
The three representations in the above proposition
are related to one another through the $\ZZ_3$ symmetry described above.
This can be presented as the following diagram:
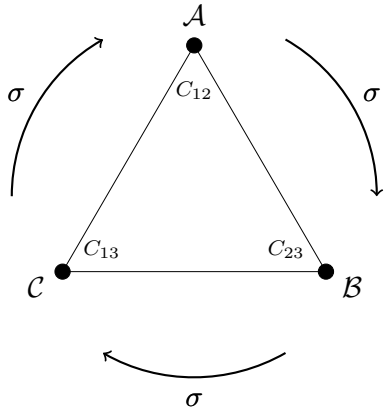
\begin{center}
\newdimen\R
\R=2cm
\begin{tikzpicture}[mystyle/.style={draw,shape=circle,fill=black}]
\draw[ultra thin, xshift=0.0\R]  {
               (90:\R) node[] {}
            -- (210:\R) node[] {}
            -- (330:\R) node[] {}
          } -- cycle (90:\R);
\draw[thick, <-] ( 1.0*1.2\R, 0.0*0.866\R) arc (000:060:1.2\R) ;
\draw[thick, <-] (-0.5*1.2\R, 1.2*0.866\R) arc (120:180:1.2\R) ;
\draw[thick, <-] (-0.5*1.2\R,-1.2*0.866\R) arc (240:300:1.2\R) ;
\foreach \a in {1,...,3} {
\draw node at (120*\a+30:1.35\R) {$\sigma$};
\draw node at (120*\a+30:1.35\R) {$\sigma$};
\draw[fill] (120*\a+90:\R) circle (0.1);
}
\draw node at (090:0.70\R) {$\scriptstyle C_{12}$};
\draw node at (210:0.70\R) {$\scriptstyle C_{13}$};
\draw node at (330:0.70\R) {$\scriptstyle C_{23}$};
\draw node at (090:1.20\R) {$\cA$};
\draw node at (210:1.20\R) {$\cC$};
\draw node at (330:1.20\R) {$\cB$};
\end{tikzpicture}
\end{center}
Each vertex corresponds to one of the representations, obtained by the set of
matrices $\cA$, $\cB$, $\cC$.
The generator $C_I$, $I=12,23,13$ shown next to each vertex indicates which of
the three generators takes a diagonal form in that particular representation.
The arrows indicate that the representations are related to one another
through the action of the $\ZZ_3$ symmetry, denoted by $\sigma$.

More precisely, we see from Remark \ref{rema:aw3Z3}
and the representation homomorphism in \eqref{eq:repaw}
that the action of $\sigma$ on the representations corresponds to
permuting the generators $\{C_{12},C_{23},C_{13}\}$ and the parameters
$\{c_1,c_2,c_3\}$ cyclically.
Starting from the expressions for the $\cA$ matrices \eqref{eq:awA},
one obtains the expressions for the $\cB$ matrices through that cyclical
permutation. Repeating the process, one gets the $\cC$ matrices, then the $\cA$
matrices, etc.
\end{rema}
\par\vspace{1em}\noindent
A natural question to ask is whether this algebra automorphism is an
\textit{outer} or an \textit{inner}
automorphism. The answer is that it is an \textit{inner automorphism}.
The three representations $\cA$, $\cB$, $\cC$ are actually isomorphic
and can be mapped to one another through a change of basis.
This will be shown explicitly in the next subsection.

\subsection{Equivalence of the three representations $\cA$, $\cB$, $\cC$}\label{}
We begin with a proposition that makes explicit the isomorphism between the $\cA$, $\cB$,
$\cC$ representations.
\begin{prop}
The three representations of the Askey--Wilson algebra $\aw$ given by the $\cA$, $\cB$,
$\cC$ sets of matrices \eqref{eq:repaw}
are equivalent to one another:
\begin{align}\label{eq:aw3_isoreps}
\cP_{\!\!\scriptscriptstyle \cA\cB}\,\cB_{I}
=\cA_{I}\,\cP_{\!\!\scriptscriptstyle \cA\cB}\,,\qquad
\cP_{\!\!\scriptscriptstyle \cB\cC}\,\cC_{I}
=\cB_{I}\,\cP_{\!\!\scriptscriptstyle \cB\cC}\,,\qquad
\cP_{\!\!\scriptscriptstyle \cC\cA}\,\cA_{I}
=\cC_{I}\,\cP_{\!\!\scriptscriptstyle \cC\cA}\,, \qquad
I\in\{12,23,13\}\,.
\end{align}
The explicit expressions of the matrices
$\cP_{\!\!\scriptscriptstyle \cA\cB}$,
$\cP_{\!\!\scriptscriptstyle \cB\cC}$,
$\cP_{\!\!\scriptscriptstyle \cC\cA}$ are given in equations
\eqref{eq:PAB}, \eqref{eq:PBC} and \eqref{eq:PCA} respectively.
Their inverses are given in equations \eqref{eq:PBA}, \eqref{eq:PCB} and \eqref{eq:PAC}.
\end{prop}
\proof
We construct the explicit change of basis matrices in order to prove the isomorphism.

Introduce the $(N+1)\times(N+1)$ matrices
$\cP_{\!\!\scriptscriptstyle \cA\cB}$,
$\cP_{\!\!\scriptscriptstyle \cB\cC}$,
$\cP_{\!\!\scriptscriptstyle \cC\cA}$.
The entries of these matrices are given by, for $0\leq i,j \leq N$:
\begin{align}\label{eq:PAB}
(\cP_{\!\!\scriptscriptstyle \cA\cB})_{i,j}&=\frac{\xi_j(c_2,c_3)}{\omega_j(c_2c_3)}~
\fp_i(j;c_3,c_2,c_1;N)\,,\\
(\cP_{\!\!\scriptscriptstyle \cB\cC})_{i,j}&=\frac{\xi_j(c_3,c_1)}{\omega_j(c_3c_1)}~
\fp_i(j;c_1,c_3,c_2;N)\,,
\label{eq:PBC}\\
(\cP_{\!\!\scriptscriptstyle \cC\cA})_{i,j}&=\frac{\xi_j(c_1,c_2)}{\omega_j(c_1c_2)}~
\fp_i(j;c_2,c_1,c_3;N)\,,
\label{eq:PCA}
\end{align}
and the functions $\xi$ and $\omega$ have been defined previously in \eqref{eq:omega}.
These matrices are invertible with their inverses
$\cP_{\!\!\scriptscriptstyle \cB\cA}$,
$\cP_{\!\!\scriptscriptstyle \cC\cB}$,
$\cP_{\!\!\scriptscriptstyle \cA\cC}$
given by, for $0\leq i,j \leq N$:
\begin{align}\label{eq:PBA}
(\cP_{\!\!\scriptscriptstyle \cB\cA})_{i,j}&=\frac{\omega_i(c_2c_3)}{\xi_i(c_2,c_3)}~
\fp_i(j;c_1,c_2,c_3;N)\,,\\
\label{eq:PCB}
(\cP_{\!\!\scriptscriptstyle \cC\cB})_{i,j}&=\frac{\omega_i(c_3c_1)}{\xi_i(c_3,c_1)}~
\fp_i(j;c_2,c_3,c_1;N)\,,\\
\label{eq:PAC}
(\cP_{\!\!\scriptscriptstyle \cA\cC})_{i,j}&=\frac{\omega_i(c_1c_2)}{\xi_i(c_1,c_2)}~
\fp_i(j;c_3,c_1,c_2;N)\,.
\end{align}
This is easily checked; using the involution relation \eqref{eq:ortho2}, one
immediately gets
\begin{align}\label{}
\cP_{\!\!\scriptscriptstyle \cA\cB}\,\cP_{\!\!\scriptscriptstyle \cB\cA}=
\cP_{\!\!\scriptscriptstyle \cB\cC}\,\cP_{\!\!\scriptscriptstyle \cC\cB}=
\cP_{\!\!\scriptscriptstyle \cC\cA}\,\cP_{\!\!\scriptscriptstyle \cA\cC}=
\un\,.
\end{align}
We claim that these matrices provide the change of basis between the
$\cA$, $\cB$, $\cC$ representations, as written in \eqref{eq:aw3_isoreps}.
Let us first examine the matrix equation
$\cP_{\!\!\scriptscriptstyle \cA\cB}\cB_{I}=\cA_{I}\cP_{\!\!\scriptscriptstyle \cA\cB}$.
For $I=12$, equating the $(n,x)$ component of the r.h.s.~and the l.h.s.~leads to
the difference equation \eqref{eq:rec-rac4} with $c_1\leftrightarrow c_3$.
For $I=23$, proceeding similarly using the equation
$\cP_{\!\!\scriptscriptstyle \cB\cA}\cA_{I}=\cB_{I}\cP_{\!\!\scriptscriptstyle \cB\cA}$
and equating the $(n,x)$ component of the r.h.s.~and the l.h.s.~also leads to
the difference equation \eqref{eq:rec-rac4}.
Since the third generator $I=13$ is obtained in terms of the first two
(see \eqref{eq:aw3_i}), it follows that $\cP_{\!\!\scriptscriptstyle \cA\cB}$ is indeed
the change of basis matrix between the $\cA$ and $\cB$ representations introduced above.
From the $\ZZ_3$ symmetry in Remark \ref{rem:Z3}, analogous results follow immediately:
$\cP_{\!\!\scriptscriptstyle \cB\cC}$ relates representations $\cB$ and $\cC$ and
$\cP_{\!\!\scriptscriptstyle \cC\cA}$ relates representations $\cC$ and $\cA$.
Therefore, equation \eqref{eq:aw3_isoreps} holds, and this completes the proof.
\endproof

We note that these changes of basis are the key ingredient in the
proof of the Racah relation for $q$-Racah polynomials \eqref{eq:racahrel}.

\section{Representations of the rank 2 Askey--Wilson algebra $\haw$}\label{app:aw4}

The higher rank version of the Askey--Wilson algebra has been introduced and studied in
\cite{PostWalt,GW,dBdC,dBdCvdV,DeC,awn,WY}.
We recall in this appendix the defining relations of $\haw$ as presented in \cite{awn},
up to a renormalization of the generators.
Then, we provide representations for this algebra.

\subsection{Defining relations for $\haw$}\label{app:def-aw4}
The algebra $\haw$ is generated by $C_I$ where $I\in\{1,2,3,4,12,23,34,123,234,1234\}$.
For these connected sequences, the order in $I$ is irrelevant, for example
$C_{21}\equiv C_{12}$.
Let us now introduce the following set of 3-tuple:
\begin{align}
\cS= \{ (1,2,3),(2,3,4),(1,23,4),(12,3,4),(1,2,34),
        (3,2,1),(4,3,2),(4,23,1),(4,3,12),(34,2,1) \} \,.
\end{align}
For $(I,J,K) \in \cS$, one defines new elements of the algebra by
\begin{align}
C_{IK}&= -\frac{1}{\scq^2-\scq^{-2}}[C_{IJ},C_{JK}]_\scq
         +\frac{1}{\scq+\scq^{-1}}(C_{I}C_{K}+C_{J}C_{IJK}) \,.
\end{align}
These relations define $C_{13},C_{24},C_{14},C_{124},C_{134}$ but also
$C_{31},C_{42},C_{41},C_{412},C_{341}$.
Let us stress that the former are different from the latter.

The defining relations of $\haw$ are:\\
{(i)\quad for $I,J\in\{1,2,3,4,12,23,34,123,234,1234\}$:}
\begin{align}
& [C_I,C_J]=0 \qquad\text{if } I\cap J =\emptyset\quad\text{or}\quad I\subset J \,,
\intertext{(ii)\quad for $(I,J,K) \in \cS$:}
&C_{IJ} = -\frac{1}{\scq^2-\scq^{-2}}[C_{JK},C_{IK}]_\scq
          +\frac{1}{\scq+\scq^{-1}}(C_{I}C_{J}+C_{K}C_{IJK})\,,\\
\intertext{and (iii):}
&C_{14} = -\frac{1}{\scq^2-\scq^{-2}}[C_{13},C_{34}]_\scq
          +\frac{1}{\scq+\scq^{-1}}(C_{1}C_{4}+C_{3}C_{134})\,,\\
&C_{41} = -\frac{1}{\scq^2-\scq^{-2}}[C_{42},C_{12}]_\scq
          +\frac{1}{\scq+\scq^{-1}}(C_{1}C_{4}+C_{2}C_{412})\,.
\end{align}

\begin{rema}\label{rema:aw4Z5}
The rank $2$ Askey--Wilson algebra $\haw$ has a $\ZZ_5$ symmetry.
In particular, the relations of the algebra $\haw$
are unchanged under the following cyclical permutations of the generators:
\begin{align}\label{eq:rho}
\rho~:\quad
\begin{aligned}
C_{1}&\mapsto C_{4}\,,\\
C_{2}&\mapsto C_{1234}\,,\\
C_{3}&\mapsto C_{1}\,,\\
C_{4}&\mapsto C_{2}\,,\\
C_{1234}&\mapsto C_{3}\,,
\end{aligned}
\qquad
\begin{aligned}
C_{12}&\mapsto C_{123}\,,\\
C_{23}&\mapsto C_{234}\,,\\
C_{34}&\mapsto C_{12}\,,\\
C_{123}&\mapsto C_{23}\,,\\
C_{234}&\mapsto C_{34}\,,
\end{aligned}
\qquad
\begin{aligned}
C_{13}&\mapsto C_{14}\,,\\
C_{24}&\mapsto C_{341}\,,\\
C_{412}&\mapsto C_{13}\,,\\
C_{14}&\mapsto C_{24}\,,\\
C_{341}&\mapsto C_{412}\,,
\end{aligned}
\qquad
\begin{aligned}
C_{31}&\mapsto C_{41}\,,\\
C_{42}&\mapsto C_{134}\,,\\
C_{124}&\mapsto C_{31}\,,\\
C_{41}&\mapsto C_{42}\,,\\
C_{134}&\mapsto C_{123}\,,
\end{aligned}
\end{align}
and it is immediate to see that $\rho^5=1$.
Note that the $\ZZ_5$ symmetry is just a part of the full braid group symmetry of the $\haw$ algebra.
We refer the reader to \cite{awn} for more details.
\end{rema}

\subsection{Representations of the algebra $\haw$}\label{sec:aw4reps}
Define
$\tilde{N}=\binom{N+2}{2}=\#\{~(i_1,i_2)~|~i_1,i_2\in\ZZ_{\geq0},~~ i_1+i_2\leq N~\}$.
We will introduce five sets of $\tilde{N}\times\tilde{N}$ matrices,
denoted by $\sA$, $\sB$, $\sC$, $\sD$, $\sE$.
Each set of matrices leads to a representation of $\haw$;
this will be stated in the next proposition down below.

We also let $\UN$ denote the $\tilde{N}$-dimensional identity matrix.

\paragraph{The $\sA$ matrices.}
Let us introduce five $\tilde{N}\times\tilde{N}$ matrices
$\sA_{12}$, $\sA_{23}$, $\sA_{34}$, $\sA_{123}$, $\sA_{234}$.
The entries of the matrices are labeled by two pairs of nonnegative indices $(i,j)$ and
$(x,y)$ such that $0\leq i+j\leq N$ and $0\leq x+y\leq N$.
The non-zero entries of the matrices $\sA_I$, $I\in\{12,23,34,123,234\}$ are given by:
\begin{align}\label{}
\begin{aligned}
(\sA_{12})_{(i,j);(i,j)}&=\lambda(i;c_1c_2)\,,\\
(\sA_{123})_{(i,j);(i,j)}&=\lambda(j;c_4c_0)\,,\\
(\sA_{34})_{(i,j-\eps);(i,j)}&=\Phi^{\eps}(j;c_3,c_4,c_0;N-i)\,,\\
(\sA_{23})_{(i-\eps',j);(i,j)}&=
\Phi^{\eps'}(i;c_3,c_2,c_1;N-j)\,
\left(\!\frac{c_3}{c_0}\!\right)^{\!\eps'/2}\!
\frac{\xi_i(c_2,c_1)}{\xi_{i-\eps'}(c_2,c_1)}\,,\\
(\sA_{234})_{(i-\eps',j-\eps);(i,j)}&=
A^{\eps,\eps'}_{j,i}(c_3,c_4,c_0,c_2;N)\,,
\end{aligned}
\qquad\eps,\eps'\in\{0,\pm\}\,.
\end{align}
It follows that $\sA_{12}$ and $\sA_{123}$ are diagonal matrices,
$\sA_{34}$ and $\sA_{23}$ are tridiagonal and the matrix $\sA_{234}$ is nine-diagonal.
These matrices provide a representation of the rank $2$ Askey--Wilson algebra,
as will be stated in a later proposition down below.

\paragraph{The $\sB$ matrices.}
The non-zero entries of the matrices $\sB_I$, $I\in\{12,23,34,123,234\}$ are given by:
\begin{align}\label{}
\begin{aligned}
(\sB_{123})_{(i,j);(i,j)}&=\lambda(i;c_4c_0)\,,\\
(\sB_{23})_{(i,j);(i,j)}&=\lambda(j;c_2c_3)\,,\\
(\sB_{12})_{(i,j-\eps);(i,j)}&=\Phi^{\eps}(j;c_1,c_2,c_3;N-i)\,,\\
(\sB_{234})_{(i-\eps',j);(i,j)}&=
\Phi^{\eps'}(i;c_1,c_0,c_4;N-j)\,
\left(\!\frac{c_1}{c_3}\!\right)^{\!\eps'/2}\!
\frac{\xi_i(c_0,c_4)}{\xi_{i-\eps'}(c_0,c_4)}\,,\\
(\sB_{34})_{(i-\eps',j-\eps);(i,j)}&=
A^{\eps,\eps'}_{j,i}(c_1,c_2,c_3,c_0;N)\,,
\end{aligned}
\qquad\eps,\eps'\in\{0,\pm\}\,.
\end{align}

\paragraph{The $\sC$ matrices.}
The non-zero entries of the matrices $\sC_I$, $I\in\{12,23,34,123,234\}$ are given by:
\begin{align}\label{}
\begin{aligned}
(\sC_{23})_{(i,j);(i,j)}&=\lambda(i;c_2c_3)\,,\\
(\sC_{234})_{(i,j);(i,j)}&=\lambda(j;c_0c_1)\,,\\
(\sC_{123})_{(i,j-\eps);(i,j)}&=\Phi^{\eps}(j;c_4,c_0,c_1;N-i)\,,\\
(\sC_{34})_{(i-\eps',j);(i,j)}&= \Phi^{\eps'}(i;c_4,c_3,c_2;N-j)\,
\left(\!\frac{c_4}{c_1}\!\right)^{\!\eps'/2}\!
\frac{\xi_i(c_3,c_2)}{\xi_{i-\eps'}(c_3,c_2)}\,,\\
(\sC_{12})_{(i-\eps',j-\eps);(i,j)}&=
A^{\eps,\eps'}_{j,i}(c_4,c_0,c_1,c_3;N)\,,
\end{aligned}
\qquad\eps,\eps'\in\{0,\pm\}\,.
\end{align}

\paragraph{The $\sD$ matrices.}
The non-zero entries of the matrices $\sD_I$, $I\in\{12,23,34,123,234\}$ are given by:
\begin{align}\label{}
\begin{aligned}
(\sD_{234})_{(i,j);(i,j)}&=\lambda(i;c_0c_1)\,,\\
(\sD_{34})_{(i,j);(i,j)}&=\lambda(j;c_3c_4)\,,\\
(\sD_{23})_{(i,j-\eps);(i,j)}&=\Phi^{\eps}(j;c_2,c_3,c_4;N-i)\,,\\
(\sD_{12})_{(i-\eps',j);(i,j)}&=
\Phi^{\eps'}(i;c_2,c_1,c_0;N-j)\,
\left(\!\frac{c_2}{c_4}\!\right)^{\!\eps'/2}\!
\frac{\xi_i(c_1,c_0)}{\xi_{i-\eps'}(c_1,c_0)}\,,\\
(\sD_{123})_{(i-\eps',j-\eps);(i,j)}&=
A^{\eps,\eps'}_{j,i}(c_2,c_3,c_4,c_1;N)\,,
\end{aligned}
\qquad\eps,\eps'\in\{0,\pm\}\,.
\end{align}

\paragraph{The $\sE$ matrices.}
The non-zero entries of the matrices $\sE_I$, $I\in\{12,23,34,123,234\}$ are given by:
\begin{align}\label{}
\begin{aligned}
(\sE_{34})_{(i,j);(i,j)}&=\lambda(i;c_3c_4)\,,\\
(\sE_{12})_{(i,j);(i,j)}&=\lambda(j;c_1c_2)\,,\\
(\sE_{234})_{(i,j-\eps);(i,j)}&=\Phi^{\eps}(j;c_0,c_1,c_2;N-i)\,,\\
(\sE_{123})_{(i-\eps',j);(i,j)}&=
\Phi^{\eps'}(i;c_0,c_4,c_3;N-j)\,
\left(\!\frac{c_0}{c_2}\!\right)^{\!\eps'/2}\!
\frac{\xi_i(c_4,c_3)}{\xi_{i-\eps'}(c_4,c_3)}\,,\\
(\sE_{23})_{(i-\eps',j-\eps);(i,j)}&=
A^{\eps,\eps'}_{j,i}(c_0,c_1,c_2,c_4;N)\,,
\end{aligned}
\qquad\eps,\eps'\in\{0,\pm\}\,.
\end{align}
We now state more precisely how the five sets of matrices
$\sA$, $\sB$, $\sC$, $\sD$, $\sE$
lead to representations of $\haw$.

\begin{prop}\label{prop:aw4rep}
When the parameter $\scq$ of $\haw$ is equal to $\sqrt{q}$, each set of
$\sA$, $\sB$, $\sC$, $\sD$, $\sE$ matrices provides
$\tilde{N}$-dimensional representations of $\haw$.
Recall that the algebra $\haw$ is generated by $C_I$ with
$I\in\{1,2,3,4,12,23,34,123,234,1234\}$ with defining relations given in
the Appendix \ref{app:def-aw4}.
Then, the following map is an algebra homomorphism:
\begin{equation}\label{eq:repaw4}
\begin{array}{lcll}
C_i & \mapsto & (\sqrt{c_i}+\frac{1}{\sqrt{c_i}})\UN\,,
& \text{for }\ i=1,2,3,4\,,
\\[1.ex]
C_{ij} & \mapsto & \frac{1}{\sqrt{qc_ic_j}}\sX_{ij}
                  +(\sqrt{qc_ic_j}+\frac{1}{\sqrt{qc_ic_j}})\UN\,,
& \text{for }\ ij=12,23,34\,,
\\[1.ex]
C_{ijk} & \mapsto & \sqrt{q^{2N+2}c_ic_jc_k}\,\sX_{ijk}
                   +(\sqrt{q^{2N+2}c_ic_jc_k}+\frac{1}{\sqrt{q^{2N+2}c_ic_jc_k}})\UN\,,
\qquad & \text{for }\ ijk=123,234\,,
\\[1.ex]
C_{1234}& \mapsto & (\sqrt{c_0}+\frac{1}{\sqrt{c_0}})\UN\,,
\end{array}
\end{equation}
where $\sX$ is chosen as either $\sA$, $\sB$, $\sC$, $\sD$ or $\sE$.
We will refer to these five representations as the
$\sA$, $\sB$, $\sC$, $\sD$ and $\sE$ representations.
\end{prop}
\proof
The defining relations of $\haw$ can be verified by explicit matrix computations
or using symbolic manipulation software such as Mathematica or Maple.
\endproof

The conditions \eqref{eq:cond-G} imply that these representations are irreducible.
Indeed, these conditions imply that the couples of eigenvalues $\big(\lambda(i;c_1c_2),\lambda(j;c_4c_0)\big)$ of the matrices $\sA_{12}$ and $\sA_{123}$ are all different.
It allows us to construct projectors on the common eigenvector $v_{i,j}$ corresponding to this couple of eigenvalues. Moreover, due to these conditions, the matrices $\sA_{34}$ and $\sA_{23}$ have non-vanishing entries
connecting $v_{i,j}$ to $v_{i,j\pm1}$ and $v_{i\pm1,j}$ respectively. Hence, through iteration, we can connect any vector $v_{i,j}$
to any other vector $v_{k,l}$. This concludes the proof of irreducibility for the representation associated to $\sA$ matrices. The proof of irreducibility for the other representations follows the same lines.

\begin{rema}\label{rem:Z5}
The five representations in the above proposition
are related to one another through the $\ZZ_5$ symmetry described above.
This can be presented as the following diagram:
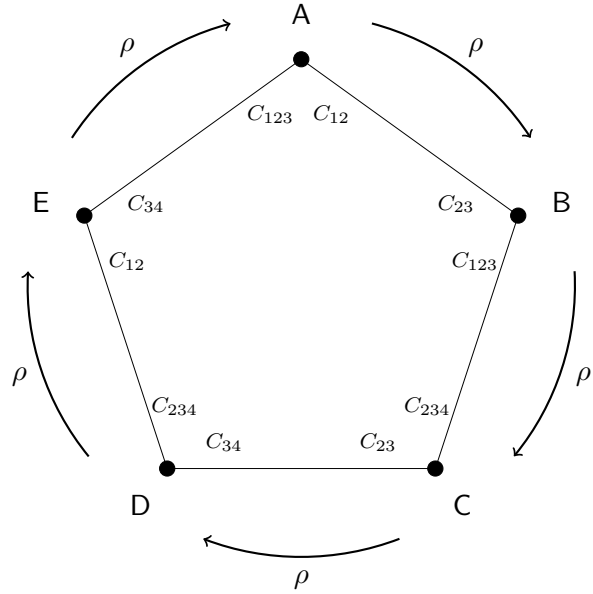
\begin{center}
\newdimen\R
\R=3cm
\begin{tikzpicture}[mystyle/.style={draw,shape=circle,fill=black}]
\draw[ultra thin, xshift=0.0\R]  {
               (018:\R) node[] {}
            -- (090:\R) node[] {}
            -- (162:\R) node[] {}
            -- (234:\R) node[] {}
            -- (306:\R) node[] {}
          } -- cycle (18:\R);
\draw[thick, <-] ( 0.838*1.2\R, 0.544*1.2\R) arc (033:075:1.2\R) ;
\draw[thick, <-] (-0.258*1.2\R, 0.965*1.2\R) arc (105:147:1.2\R) ;
\draw[thick, <-] (-0.998*1.2\R, 0.052*1.2\R) arc (177:219:1.2\R) ;
\draw[thick, <-] (-0.358*1.2\R,-0.933*1.2\R) arc (249:291:1.2\R) ;
\draw[thick, <-] ( 0.777*1.2\R,-0.629*1.2\R) arc (321:363:1.2\R) ;
\foreach \a in {1,...,5} {
\draw node at (72*\a+54:1.3\R) {$\rho$};
\draw[fill] (72*\a+90:\R) circle (0.1);
}

\draw node at (080:0.77\R) {$\scriptstyle C_{12}$};
\draw node at (100:0.77\R) {$\scriptstyle C_{123}$};
\draw node at (152:0.77\R) {$\scriptstyle C_{34}$};
\draw node at (172:0.77\R) {$\scriptstyle C_{12}$};
\draw node at (224:0.77\R) {$\scriptstyle C_{234}$};
\draw node at (244:0.77\R) {$\scriptstyle C_{34}$};
\draw node at (296:0.77\R) {$\scriptstyle C_{23}$};
\draw node at (316:0.77\R) {$\scriptstyle C_{234}$};
\draw node at (008:0.77\R) {$\scriptstyle C_{123}$};
\draw node at (028:0.77\R) {$\scriptstyle C_{23}$};

\draw node at (090:1.20\R) {$\sA$};
\draw node at (162:1.20\R) {$\sE$};
\draw node at (234:1.20\R) {$\sD$};
\draw node at (306:1.20\R) {$\sC$};
\draw node at (018:1.20\R) {$\sB$};
\end{tikzpicture}
\end{center}
Each vertex corresponds to one of the representations, obtained by the set of
matrices $\sA$, $\sB$, $\sC$, $\sD$, $\sE$.
The two generators $C_I$, $I\in\{12,23,34,123,234\}$ shown next to each vertex indicate
which two of the five generators take a diagonal form in that particular representation.
The arrows indicate that the representations are related to one another through the action
of the $\ZZ_5$ symmetry, denoted by $\rho$.

More precisely, we see from Remark \ref{rema:aw4Z5}
and the representation homomorphism in \eqref{eq:repaw4}
that the action of $\rho$ on the representations corresponds to
permuting the generators $\{C_{12},C_{123},C_{23},C_{234},C_{34}\}$ and the parameters
$\{c_1,c_4,c_2,c_0,c_3\}$ cyclically.
Starting from the expressions for the $\sA$ matrices \eqref{eq:awA},
one obtains the expressions for the $\sB$ matrices through that cyclical
permutation. Repeating the process, one gets the $\sC$ matrices, then the $\sD$
matrices, the $\sE$ matrices, the $\sA$ matrices, etc.
\end{rema}
\par\vspace{1em}\noindent
Again, it is natural to ask whether this algebra automorphism is an
\textit{outer} or an \textit{inner}
automorphism. The answer here is again that it is an \textit{inner automorphism}.
The five representations $\sA$, $\sB$, $\sC$, $\sD$, $\sE$ are actually isomorphic
and can be mapped to one another through a change of basis.
This will be shown explicitly in the next subsection.

For completeness, we now provide the expressions for other generators of the
$\haw$ algebra in the $\sA$ representation below. The following are
tridiagonal matrices:
\begin{align}\label{}
\begin{aligned}
(\sA_{13})_{(i-\eps',j);(i,j)} &= \Phi^{\eps'}(i;c_3,c_1,c_2;N-j)\,
\left(\!\frac{c_3}{c_0}\!\right)^{\!\eps'/2}\!
\frac{\omega_{i-\eps'}(c_2c_1)}{\omega_{i}(c_2c_1)}\,,\\
(\sA_{31})_{(i-\eps',j);(i,j)} &= \Phi^{\eps'}(i;c_3,c_1,c_2;N-j)\,
\left(\!\frac{c_3}{c_0}\!\right)^{\!\eps'/2}\!
\frac{\omega_{i}(c_2c_1)}{\omega_{i-\eps'}(c_2c_1)}\,,\\
(\sA_{124})_{(i,j-\eps);(i,j)}&= \Phi^{\eps}(j;c_3,c_0,c_4;N-i)\,
\frac{\xi_{j-\eps}(c_4,c_0)\omega_{j-\eps}(c_4c_0)}
     {\xi_{j}(c_4,c_0)\omega_{j}(c_4c_0)}\,,\\
(\sA_{412})_{(i,j-\eps);(i,j)}&= \Phi^{\eps}(j;c_3,c_0,c_4;N-i)\,
\frac{\xi_{j}(c_0,c_4)\omega_{j}(c_0c_4)}
     {\xi_{j-\eps}(c_0,c_4)\omega_{j-\eps}(c_0c_4)}\,,
\end{aligned}
\end{align}
and the following are nine-diagonal matrices:
\begin{align}\label{}
\begin{aligned}
(\sA_{24})_{(i-\eps',j-\eps);(i,j)} &= A^{\eps,\eps'}_{j,i}(c_3,c_0,c_4,c_1;N)\,
\frac{\xi_{j-\eps}(c_4,c_0)\omega_{j-\eps}(c_4c_0)}
     {\xi_{j}(c_4,c_0)\omega_{j}(c_4c_0)}\!
\left(\!\frac{c_4}{c_0}\!\right)^{\!\eps'/2}\!\!
\frac{\xi_{i}(c_2,c_1)\omega_{i}(c_2c_1)}
     {\xi_{i-\eps'}(c_2,c_1)\omega_{i-\eps'}(c_2c_1)}\,,\\
(\sA_{42})_{(i-\eps',j-\eps);(i,j)} &= A^{\eps,\eps'}_{j,i}(c_3,c_0,c_4,c_1;N)\,
\frac{\xi_{j}(c_0,c_4)\omega_{j}(c_0c_4)}
     {\xi_{j-\eps}(c_0,c_4)\omega_{j-\eps}(c_0c_4)}\!
\left(\!\frac{c_4}{c_0}\!\right)^{\!\eps'/2}\!\!
\frac{\xi_{i-\eps'}(c_1,c_2)\omega_{i-\eps'}(c_1c_2)}
     {\xi_{i}(c_1,c_2)\omega_{i}(c_1c_2)}\,,\\
(\sA_{14})_{(i-\eps',j-\eps);(i,j)} &= A^{\eps,\eps'}_{j,i}(c_3,c_0,c_4,c_2;N)\,
\frac{\xi_{j-\eps}(c_4,c_0)\omega_{j-\eps}(c_4c_0)}
     {\xi_{j}(c_4,c_0)\omega_{j}(c_4c_0)}\!
\left(\!\frac{c_4}{c_0}\!\right)^{\!\eps'/2}\,,\\
(\sA_{41})_{(i-\eps',j-\eps);(i,j)} &= A^{\eps,\eps'}_{j,i}(c_3,c_0,c_4,c_2;N)\,
\frac{\xi_{j}(c_0,c_4)\omega_{j}(c_0c_4)}
     {\xi_{j-\eps}(c_0,c_4)\omega_{j-\eps}(c_0c_4)}\!
\left(\!\frac{c_4}{c_0}\!\right)^{\!\eps'/2}\,,\\
(\sA_{134})_{(i-\eps',j-\eps);(i,j)} &= A^{\eps,\eps'}_{j,i}(c_3,c_4,c_0,c_1;N)\,
\frac{\xi_{i-\eps'}(c_1,c_2)\omega_{i-\eps'}(c_1c_2)}
     {\xi_{i}(c_1,c_2)\omega_{i}(c_1c_2)}\,,\\
(\sA_{341})_{(i-\eps',j-\eps);(i,j)} &= A^{\eps,\eps'}_{j,i}(c_3,c_4,c_0,c_1;N)\,
\frac{\xi_{i}(c_2,c_1)\omega_{i}(c_2c_1)}
     {\xi_{i-\eps'}(c_2,c_1)\omega_{i-\eps'}(c_2c_1)}\,,
\end{aligned}
\end{align}
with $\eps,\eps'\in\{0,\pm\}$.
The expressions for the matrices in the other bases
$\sB$, $\sC$, $\sD$, $\sE$
are directly obtained from those in the $\sA$ basis
by making use of the $\ZZ_5$ symmetry, see remark \ref{rem:Z5}.

For the five representations given in the above proposition, the generators associated to
non-connected sequences (see the definition in Section \ref{app:def-aw4}) are then given
as follows, with $\sX\in\{\sA,\sB,\sC,\sD,\sE\}$:
\begin{equation}\label{eq:repaw4_sup}
\begin{array}{lcll}
C_{ij} & \mapsto & \frac{1}{\sqrt{qc_ic_j}}\sX_{ij}
                  +(\sqrt{qc_ic_j}+\frac{1}{\sqrt{qc_ic_j}})\UN\,,
& \text{for }\ ij=13,24,14,31,42,41\,,
\\[1.ex]
C_{ijk} & \mapsto & \sqrt{q^{2N+2}c_ic_jc_k}\,\sX_{ijk}
                   +(\sqrt{q^{2N+2}c_ic_jc_k}+\frac{1}{\sqrt{q^{2N+2}c_ic_jc_k}})\UN\,,
\quad & \text{for }\ ijk=124,134,412,341\,.
\end{array}
\end{equation}
Once again, the algebraic relations can be verified by explicit matrix computations
or using symbolic manipulation software such as Mathematica or Maple.

\subsection{Equivalence of the five representations $\sA$, $\sB$, $\sC$, $\sD$, $\sE$}
\label{ssec:5bases}
We begin with a proposition that makes the isomorphism between the
$\sA$, $\sB$, $\sC$, $\sD$, $\sE$
representations explicit.
\begin{prop}\label{prop:iso_aw4_reps}
The five representations of the rank $2$ Askey--Wilson algebra $\haw$ given by the
$\sA$, $\sB$, $\sC$, $\sD$, $\sE$ matrices are equivalent to one another:
\begin{gather}\label{eq:aw4_isoreps}
\begin{gathered}
\sP_{\!\!\scriptscriptstyle \sA\sB}\,\sB_{I}
=\sA_{I}\,\sP_{\!\!\scriptscriptstyle \sA\sB}\,,\qquad
\sP_{\!\!\scriptscriptstyle \sB\sC}\,\sC_{I}
=\sB_{I}\,\sP_{\!\!\scriptscriptstyle \sB\sC}\,,\qquad
\sP_{\!\!\scriptscriptstyle \sC\sD}\,\sD_{I}
=\sC_{I}\,\sP_{\!\!\scriptscriptstyle \sC\sD}\,,\\
\sP_{\!\!\scriptscriptstyle \sD\sE}\,\sE_{I}
=\sD_{I}\,\sP_{\!\!\scriptscriptstyle \sD\sE}\,,\qquad
\sP_{\!\!\scriptscriptstyle \sE\sA}\,\sA_{I}
=\sE_{I}\,\sP_{\!\!\scriptscriptstyle \sE\sA}\,,\qquad
\end{gathered}\qquad
I\in\{12,23,34,123,234\}\,.
\end{gather}
The explicit expressions of the change of basis matrices
$\sP_{\!\!\scriptscriptstyle \sA\sB}$,
$\sP_{\!\!\scriptscriptstyle \sB\sC}$,
$\sP_{\!\!\scriptscriptstyle \sC\sD}$,
$\sP_{\!\!\scriptscriptstyle \sD\sE}$,
$\sP_{\!\!\scriptscriptstyle \sE\sF}$
and their inverses are given in the proof below.
\end{prop}
\proof
We construct the explicit change of basis matrices in order to prove the isomorphism.

Introduce the $\widetilde{N}\times\widetilde{N}$ matrices
$\sP_{\!\!\scriptscriptstyle \sA\sB}$,
$\sP_{\!\!\scriptscriptstyle \sB\sC}$,
$\sP_{\!\!\scriptscriptstyle \sC\sD}$,
$\sP_{\!\!\scriptscriptstyle \sD\sE}$,
$\sP_{\!\!\scriptscriptstyle \sE\sF}$.
The entries of these matrices are given as follows,
for $0\leq i_1+i_2\leq N$ and $0\leq j_1+j_2\leq N$:
\begin{align}\label{}
(\sP_{\!\!\scriptscriptstyle \sA\sB})_{(i_1,i_2);(j_1,j_2)}&=
\delta_{i_2,j_1} \left(\!\frac{c_0}{c_3}\!\right)^{\!\!\nicefrac{i_1}{2}}
\xi_{i_1}(c_1,c_2)~ \fp_{i_1}(j_2;c_3,c_2,c_1;N-i_2)\,,\\
(\sP_{\!\!\scriptscriptstyle \sB\sC})_{(i_1,i_2);(j_1,j_2)}&=
\delta_{i_2,j_1} \left(\!\frac{c_3}{c_1}\!\right)^{\!\!\nicefrac{i_1}{2}}
\xi_{i_1}(c_4,c_0)~ \fp_{i_1}(j_2;c_1,c_0,c_4;N-i_2)\,,\\
(\sP_{\!\!\scriptscriptstyle \sC\sD})_{(i_1,i_2);(j_1,j_2)}&=
\delta_{i_2,j_1} \left(\!\frac{c_1}{c_4}\!\right)^{\!\!\nicefrac{i_1}{2}}
\xi_{i_1}(c_2,c_3)~ \fp_{i_1}(j_2;c_4,c_3,c_2;N-i_2)\,,\\
(\sP_{\!\!\scriptscriptstyle \sD\sE})_{(i_1,i_2);(j_1,j_2)}&=
\delta_{i_2,j_1} \left(\!\frac{c_4}{c_2}\!\right)^{\!\!\nicefrac{i_1}{2}}
\xi_{i_1}(c_0,c_1)~ \fp_{i_1}(j_2;c_2,c_1,c_0;N-i_2)\,,\\
(\sP_{\!\!\scriptscriptstyle \sE\sA})_{(i_1,i_2);(j_1,j_2)}&=
\delta_{i_2,j_1} \left(\!\frac{c_2}{c_0}\!\right)^{\!\!\nicefrac{i_1}{2}}
\xi_{i_1}(c_3,c_4)~ \fp_{i_1}(j_2;c_0,c_4,c_3;N-i_2)\,,
\end{align}
and the function $\xi$ has been defined previously in \eqref{eq:omega}.
The inverses of the above matrices, namely
$\sP_{\!\!\scriptscriptstyle \sB\sA}$,
$\sP_{\!\!\scriptscriptstyle \sC\sB}$,
$\sP_{\!\!\scriptscriptstyle \sD\sC}$,
$\sP_{\!\!\scriptscriptstyle \sE\sD}$,
$\sP_{\!\!\scriptscriptstyle \sA\sE}$,
have entries given as follows,
for $0\leq i_1+i_2\leq N$ and $0\leq j_1+j_2\leq N$:
\begin{align}\label{}
(\sP_{\!\!\scriptscriptstyle \sB\sA})_{(i_1,i_2);(j_1,j_2)}&=
\delta_{i_1,j_2} \left(\!\frac{c_3}{c_0}\!\right)^{\!\!\nicefrac{j_1}{2}}
{\xi_{j_1}(c_2,c_1)}~ \fp_{i_2}(j_1;c_1,c_2,c_3;N-j_2)\,,\\
(\sP_{\!\!\scriptscriptstyle \sC\sB})_{(i_1,i_2);(j_1,j_2)}&=
\delta_{i_1,j_2} \left(\!\frac{c_1}{c_3}\!\right)^{\!\!\nicefrac{j_1}{2}}
{\xi_{j_1}(c_0,c_4)}~ \fp_{i_2}(j_1;c_4,c_0,c_1;N-j_2)\,,\\
(\sP_{\!\!\scriptscriptstyle \sD\sC})_{(i_1,i_2);(j_1,j_2)}&=
\delta_{i_1,j_2} \left(\!\frac{c_4}{c_1}\!\right)^{\!\!\nicefrac{j_1}{2}}
{\xi_{j_1}(c_3,c_2)}~ \fp_{i_2}(j_1;c_2,c_3,c_4;N-j_2)\,,\\
(\sP_{\!\!\scriptscriptstyle \sE\sD})_{(i_1,i_2);(j_1,j_2)}&=
\delta_{i_1,j_2} \left(\!\frac{c_2}{c_4}\!\right)^{\!\!\nicefrac{j_1}{2}}
{\xi_{j_1}(c_1,c_0)}~ \fp_{i_2}(j_1;c_0,c_1,c_2;N-j_2)\,,\\
(\sP_{\!\!\scriptscriptstyle \sA\sE})_{(i_1,i_2);(j_1,j_2)}&=
\delta_{i_1,j_2} \left(\!\frac{c_0}{c_2}\!\right)^{\!\!\nicefrac{j_1}{2}}
{\xi_{j_1}(c_4,c_3)}~ \fp_{i_2}(j_1;c_3,c_4,c_0;N-j_2)\,.
\end{align}
This is easily checked; using the involution relation \eqref{eq:ortho2}, one gets
\begin{align}\label{}
\sP_{\!\!\scriptscriptstyle \sA\sB}\,\sP_{\!\!\scriptscriptstyle \sB\sA}=
\sP_{\!\!\scriptscriptstyle \sB\sC}\,\sP_{\!\!\scriptscriptstyle \sC\sB}=
\sP_{\!\!\scriptscriptstyle \sC\sD}\,\sP_{\!\!\scriptscriptstyle \sD\sC}=
\sP_{\!\!\scriptscriptstyle \sD\sE}\,\sP_{\!\!\scriptscriptstyle \sE\sD}=
\sP_{\!\!\scriptscriptstyle \sE\sA}\,\sP_{\!\!\scriptscriptstyle \sA\sE}=
\UN\,.
\end{align}
We claim that these matrices provide the change of basis between the
$\sA$, $\sB$, $\sC$, $\sD$, $\sE$
representations as written in \eqref{eq:aw4_isoreps}.

First examine the change of basis between the $\sA$ and $\sB$ bases.

For $I=123$, look at the change of basis equation \eqref{eq:aw4_isoreps}
$\sA_{123}\,\sP_{\!\!\scriptscriptstyle \sA\sB}
=\sP_{\!\!\scriptscriptstyle \sA\sB}\,\sB_{123}$.
The entries in the l.h.s.~and r.h.s.~are identical so the equation is satisfied.

For $I=12$, again look at the components of the equation
$\sA_{123}\,\sP_{\!\!\scriptscriptstyle \sA\sB}
=\sP_{\!\!\scriptscriptstyle \sA\sB}\,\sB_{123}$.
Equating the $(n,k);(k,x)$ component of the l.h.s.~and r.h.s.~leads to
the difference equation \eqref{eq:rec-rac4} with $c_1\leftrightarrow c_3$ and
$N\mapsto N-k$
(for $0\leq n,x\leq N-k$) hence this equation holds.

Similarly, for $I=23$, rewrite the change of basis equation \eqref{eq:aw4_isoreps} as
$\sB_{23}\,\sP_{\!\!\scriptscriptstyle \sB\sA}
=\sP_{\!\!\scriptscriptstyle \sB\sA}\,\sA_{23}$.
Equating the $(k,n);(x,k)$ component of the l.h.s.~and r.h.s.~leads to
the difference equation \eqref{eq:rec-rac4} with $N\mapsto N-k$
(for $0\leq n,x\leq N-k$) hence this equation also holds.

For $I=34$, rewrite the change of basis equation \eqref{eq:aw4_isoreps} as
$\sA_{34}\,\sP_{\!\!\scriptscriptstyle \sA\sB}
=\sP_{\!\!\scriptscriptstyle \sA\sB}\,\sB_{34}$.
Upon comparing the $(n,k);(k,x)$ components of the l.h.s.~and the r.h.s.~one
checks that they are indeed equal by virtue of the difference equation
\eqref{eq:rec-rac4}.
Simultaneously, comparing the $(n,k-\eps);(k,x)$ components of the l.h.s.~and the
r.h.s.~for $\eps=\pm$, one confirms that they are equal by virtue of
the contiguity difference equations \eqref{eq:cont-rec-rac2}
involving $\lambda_{\eps}$ respectively.
Thus, the equation also holds in this case.

Finally, an analysis along the same lines as what was done for $I=34$ can be repeated for
$I=234$. Rewriting the change of basis equation \eqref{eq:aw4_isoreps} as
$\sB_{234}\,\sP_{\!\!\scriptscriptstyle \sB\sA}
=\sP_{\!\!\scriptscriptstyle \sB\sA}\,\sA_{234}$,
one finds that this equation also holds.

With this, we have proven the validity of the change of basis between the $\sA$ and
$\sB$ bases. The remaining changes of basis follow immediately by the $\ZZ_5$ symmetry.
\endproof

Again let us remark that these changes of basis are the key ingredient in the proof of
the Biedenharn--Elliot relation for $q$-Racah polynomials \eqref{eq:biedenrel}.

\end{document}